\documentclass[epsfig,12pt]{article}
\usepackage{graphicx}
\usepackage{epstopdf}

\usepackage{array}
\usepackage{amsmath}
\usepackage{amssymb}

\newcommand{\beq}{\begin{equation}}   
\newcommand{\eeq}{\end{equation}}
\newcommand{\beqn}{\begin{eqnarray}}   
\newcommand{\eeqn}{\end{eqnarray}}

\usepackage{slashed}
\begin{document}
\unitlength = 1mm

\def\de{\partial}
\def\Tr{ \hbox{\rm Tr}}
\def\const{\hbox {\rm const.}}  
\def\o{\over}
\def\im{\hbox{\rm Im}}
\def\re{\hbox{\rm Re}}
\def\bra{\langle}\def\ket{\rangle}
\def\Arg{\hbox {\rm Arg}}
\def\Re{\hbox {\rm Re}}
\def\Im{\hbox {\rm Im}}
\def\diag{\hbox{\rm diag}}


\def\QATOPD#1#2#3#4{{#3 \atopwithdelims#1#2 #4}}
\def\stackunder#1#2{\mathrel{\mathop{#2}\limits_{#1}}}
\def\stackreb#1#2{\mathrel{\mathop{#2}\limits_{#1}}}
\def\Tr{{\rm Tr}}
\def\res{{\rm res}}
\def\Bf#1{\mbox{\boldmath $#1$}}
\def\balpha{{\Bf\alpha}}
\def\bbeta{{\Bf\beta}}
\def\bgamma{{\Bf\gamma}}
\def\bnu{{\Bf\nu}}
\def\bmu{{\Bf\mu}}
\def\bphi{{\Bf\phi}}
\def\bPhi{{\Bf\Phi}}
\def\bomega{{\Bf\omega}}
\def\blambda{{\Bf\lambda}}
\def\brho{{\Bf\rho}}
\def\bsigma{{\bfit\sigma}}
\def\bxi{{\Bf\xi}}
\def\bbeta{{\Bf\eta}}
\def\d{\partial}
\def\der#1#2{\frac{\d{#1}}{\d{#2}}}
\def\Im{{\rm Im}}
\def\Re{{\rm Re}}
\def\rank{{\rm rank}}
\def\diag{{\rm diag}}
\def\2{{1\over 2}}
\def\ntwo{${\mathcal N}=2\;$}
\def\nfour{${\mathcal N}=4\;$}
\def\none{${\mathcal N}=1\;$}
\def\ntwot{${\mathcal N}=(2,2)\;$}
\def\ntwoo{${\mathcal N}=(0,2)\;$}
\def\x{\stackrel{\otimes}{,}}

\newcommand{\cpn}{CP$(N-1)\;$}
\newcommand{\wcpn}{wCP$_{N,\tilde{N}}(N_f-1)\;$}
\newcommand{\wcpd}{wCP$_{\tilde{N},N}(N_f-1)\;$}
\newcommand{\vp}{\varphi}
\newcommand{\pt}{\partial}
\newcommand{\ve}{\varepsilon}
\renewcommand{\theequation}{\thesection.\arabic{equation}}

\newcommand{\sun}{SU$(N)\;$}

\setcounter{footnote}0

\vfill

\begin{titlepage}

\begin{flushright}
FTPI-MINN-15/18, UMN-TH-3431/15\\
May 28, 2015
\end{flushright}

\vspace{1mm}

\begin{center}
{  \Large \bf  
 Non-Abelian String of a Finite Length
}

\vspace{5mm}

 {\large \bf  S.~Monin$^{\,a}$,  M.~Shifman$^{\,a}$ and \bf A.~Yung$^{\,\,a,b,c}$}
\end {center}

\begin{center}

$^a${\it  William I. Fine Theoretical Physics Institute,
University of Minnesota,
Minneapolis, MN 55455, USA}\\
$^{b}${\it National Research Center ``Kurchatov Institute'', 
Petersburg Nuclear Physics Institute, Gatchina, St. Petersburg
188300, Russia}\\
$^{c}${\it  St. Petersburg State University,
 St. Petersburg 198504, Russia}
\end{center}

\vspace{1mm}

\begin{center}
{\large\bf Abstract}
\end{center}

We consider world-sheet theories for non-Abelian strings assuming compactification
on a cylinder with a finite circumference $L$ and periodic boundary conditions. 
 The dynamics of the orientational modes is
described by two-dimensional CP$(N-1)$ model. 
 We analyze both non-supersymmetric (bosonic) model and ${\mathcal N}=(2,2)$ supersymmetric CP$(N-1)$ 
 emerging in the case of 
1/2-BPS saturated strings in \ntwo supersymmetric QCD with $N_f=N$. 
The non-supersymmetric case was studied previously; technically our results agree with those obtained previously, although our interpretation is totally different. In the large-$N$ limit we detect a phase transition at $L\sim \Lambda_{\rm CP}^{-1}$ (which is expected to become a rapid crossover at finite $N$).
If at large $L$ the CP$(N-1)$ model develops a mass gap and 
is in the Coulomb/confinement 
phase, with exponentially suppressed finite-$L$ effects, at small $L$ it is in the deconfinement
phase, and the orientational modes contribute to the L\"usher term. The latter becomes dependent on the rank of 
the bulk gauge group. 

In the supersymmetric CP$(N-1)$ models at finite $L$ we find a large-$N$ 
solution which was not known previously.  We observe a single phase independently of the value 
of $L\Lambda_{\rm CP}$. For any value of this parameter  a mass gap develops and supersymmetry
remains unbroken. So does the $SU(N)$ symmetry of the target space. 
The mass gap turns out to be independent of the  string length. The L\"uscher term is absent due to supersymmetry.

\vspace{2cm}

\end{titlepage}

 \newpage

\tableofcontents

\newpage

\section{Introduction}
\label{intro}
\setcounter{equation}{0}

Recently there was a considerable progress in studies of  long confining strings,
 see   \cite{AharonyKomar}. The energy of the
 Abrikosov-Nielsen-Olesen (ANO) closed string \cite{ANO} in the Abelian-Higgs model as a function of the
string length $L$ (in the large-$L$ limit) can be written as 
\beq
E(L) = TL - \frac{\gamma}{L} + \frac{c_3}{TL^3} +\cdots,
\label{ANOexp}
\eeq
where $T$ is the string tension and ellipses stand for  terms of the higher order in $1/L$. This $1/L$
expansion 
is determined by the low-energy effective two-dimensional theory on the string world-sheet.
For the ANO
string the world-sheet theory is given by the
Nambu-Goto action plus higher derivative corrections. It is plausible to assume that a
a similar structure applies to QCD confining strings. Recently a significant progress occurred
 in measuring
the spectrum of long confining QCD strings  in lattice simulations, see, for example, 
\cite{lattice}.

The $1/L$ term in (\ref{ANOexp}) is referred to as the L\"uscher term \cite{L}. The coefficient $\gamma$ is
 universal. Its value is determined by the number of massless
(light) degrees of freedom on the string world-sheet. The Abelian strings possess only two massless 
excitations due to two translational zero modes;  the L\"uscher term is, 
correspondingly, $\displaystyle\gamma={\pi}/{3}$.

In this paper we will study the energy of a finite-$L$ closed {\em non}-Abelian string assuming that $L$ is much larger than the string transverse size.

The main feature of the non-Abelian strings is the occurrence of extra (quasi)moduli:
orienational zero modes associated with their color flux rotation in the internal space. Dynamics 
of these orientational moduli is described by two-dimensional CP$(N-1)$ model on the 
string world-sheet. If the bulk theory supporting such string vortices is supersymmetric,\footnote{In the simplest version non-Abelian vortex strings are 
supported in  gauge theories with the U$(N)$ gauge group and $N_f=N$ flavors of quarks.} the world-sheet 
CP$(N-1)$ model will have various degrees of supersymmetry.
Non-Abelian strings were 
first found in \ntwo supersymmetric gauge theories \cite{HT1,ABEKY,SYmon,HT2}. Later
this construction was generalized to a wide class of non-Abelian gauge theories, both supersymmetric
and non-supersymmetric, see \cite{Trev,Jrev,SYrev,Trev2}. The L\"usher term for non-supersymmetric non-Abelian strings was previously discussed in \cite{ShYu}.

Our current task is broader: we want to study the $L$ dependence of $E(L)$ for all values of $L$, 
large and small
(see below), taking account of 
the orientational moduli that are described by two-dimensional CP$(N-1)$ model. The latter is
asymptotically free and develops  its own dynamical scale $\Lambda_{\rm CP}$. This modifies the expansion
in (\ref{ANOexp}). Assuming that 
\beq
\Lambda_{\rm CP}\ll \sqrt{T}
\label{largeT}
\eeq 
we can write
\beq
E(L) = TL + \frac{f(\Lambda_{CP}L)}{L} + O\left(\frac{1}{TL^3}\right).
\label{NAexp}
\eeq

Below we will present a detailed calculation of the string energy for strings with
\beqn
L\gg 1/\sqrt{T}\,.
\label{llarge}
\eeqn
 For these values  of $L$ higher derivative corrections to the effective
world-sheet theory can be ignored, and we use  CP$(N-1)$-based description to calculate the function
$f(\Lambda_{CP}L)$ (which is already known \cite{ShYu} in the limits $L\gg \Lambda_{\rm CP}^{-1}\,$ and $L \ll  \Lambda_{\rm CP}^{-1} $). To 
solve the CP$(N-1)$ model we use the large-$N$ approximation \cite{W}.
Given the constraint (\ref{llarge}) which is also assumed, we call the string ``large" if $L\gg \Lambda_{\rm CP}^{-1}\,$, and ``small" otherwise. 

Now, when we have two free parameters in the problem under consideration, $N$ and $L$, and both can be large,
the ordering of taking limits is of paramount importance and  a source of a number of paradoxes.
We will {\em always} take first the limit $N\to\infty$. In this limit the number of dynamical degrees of freedom is infinite (even in the quantum-mechanical limit $L\to 0$) and, moreover, all interactions die off. This makes possible phase transitions. 

 For non-supersymmetric case we find a phase transition 
in the CP$(N-1)$ model on the string world-sheet. Its origin is intuitively clear:
at large $L$ the theory is strongly coupled while at small $L$ it is weakly coupled and its behavior should be close
to that given by perturbation theory. Correspondingly, 
at large string length this theory develops a mass gap and 
is in the Coulomb/confinement 
phase. Finite-length effects coming from orientational moduli are exponentially suppressed. 
We find that at  $L \gg\Lambda_{\rm CP}$
\beq
f(\Lambda_{CP}L) =-\frac{\pi}{3} - N\sqrt{\frac{2}{\pi}}\,\sqrt{\Lambda_{\rm CP}L}\,e^{-\Lambda_{\rm CP}L}
+ \cdots ,
\label{flargeL}
\eeq
where the first term is the conventional  L\"uscher term coming from the translational moduli.

At small length the  CP$(N-1)$ model is in the deconfinement
phase. Massless orientational moduli contribute to the L\"uscher term
 which becomes dependent on the rank of  the bulk gauge group. At  $\sqrt{T} \ll L \ll\Lambda_{\rm CP}$ we find
that 
\beq
f(\Lambda_{\rm CP}L) =- N\,\frac{\pi }{3}.
\label{fsmallL}
\eeq

The asymptotic values of  the L\"uscher coefficient $\gamma $
associated with the limits of large and small $L$ in (\ref{flargeL}) and (\ref{fsmallL}), 
respectively, were reported earlier in  \cite{ShYu} for the open string. Here we confirm these results and derive 
$f(\Lambda_{\rm CP}L)$ for the closed sting. In other words, we impose periodic boundary conditions (on the boson
and fermion fields in the case of supersymmetric model, see below).

If $N$ is large but finite we expect that the phase transition becomes a rapid crossover. We do not expect
strictly massless states to appear in the small-$L$ domain at finite $N$.

Next, we study supersymmetric case considering BPS-saturated non-Abelian string in four-dimensional 
\ntwo SQCD. In this case the world-sheet theory for orientational modes is \ntwot supersymmetric
 CP$(N-1)$ model. Solving this theory in the large-$N$ limit we
 find a {\em single} phase with unbroken supersymmetry and a  mass gap.
 The mass gap
turns out to be independent of the  string length.
  The chiral $Z_{2N}$ symmetry is broken down 
to $Z_2$,  in much the same way as for infinitely long string. The photon field acquires a mass term, 
and no Coulomb/confining
potential is generated. Instead, the theory has $N$  degenerate vacua representing $N$ elementary strings. 
The  L\"uscher term vanishes due to the boson-fermion
cancellation. 

Thus, the dynamical $L$-behavior of non-Abelian strings, with or without supersymmetry, is drastically different in the large-$N$ solution.

\vspace{1mm}

As was mentioned, in both cases we impose periodic boundary conditions
on the spacial interval of length $L$.
In the non-supersymmetric case this is equivalent to endowing the string under consideration with 
temperature $\beta^{-1}$,
\beq
\beta =L\,.
\eeq

Such strings were considered previously, see e.g. \cite{Aff1,Laz,Actor}. Our results differ from those of
\cite{Aff1,Laz,Actor} partly in interpretation and partly in essence. 

The paper is organized as follows. In Sections \ref{nsnas} and \ref{cpn1mazt} we briefly review non-supersymmetric bulk theory supporting
non-Abelian strings and the large-$N$ solution of the CP$(N-1)$ model at $L\to\infty$ \cite{W}, respectively.
 In Sec. \ref{seclargeL} we
use the large-$N$ method to study non-Abelian strings of finite length and, in particular, describe the
Coulomb/confinement phase. Section \ref{secdeconfinement}  is devoted to  deconfinement phase. In Secs. \ref{susyrev} and  \ref{susy}, central in our analysis, we deal with supersymmetric ${\mathcal N}=(2,2)$ string. 
In Sec. \ref{phma} we calculate the photon mass on the world-sheet of the supersymmetric string under consideration as a function of
$L$.
Sec. \ref{conclusions} summarizes our conclusions. Appendices contain details of our calculations.

\section{Non-supersymmetric non-Abelian strings}
\label{nsnas}
\setcounter{equation}{0}

In this section we briefly review the simplest four-dimensional non-supersym\-metric model 
supporting non-Abelian strings
 \cite{GSY1},
 give a topological argument for their stability and outline  the effective low-energy 
theory on the world-sheet. 

The model  suggested in \cite{GSY1} is a bosonic part of \ntwo supersymmetric QCD, see \cite{SYrev}
for a review. The gauge group of the 
theory is $SU(N)\times U(1)$. The matter sector of the model consists of $N_f=N$ flavors of complex 
scalar fields (squarks) charged with respect to $U(1)$,  each in the fundamental representation of $SU(N)$. 
The action of the model is 

\beqn
S&=&\int d^4x\Bigl [-\frac{1}{4g_2^2}\left(F^a_{\mu\nu}\right)^2-\frac{1}{4g_1^2}\left(F_{\mu\nu}\right)^2\nonumber \\[3mm]
&+&|\nabla^\mu\varphi^A|^2+\frac{g_2^2}{2}\left(\bar\varphi_AT^a\varphi^A\right)^2+\frac{g_1^2}{8}\left(|\varphi^A|^2-N\xi\right)^2\Bigr] \,,
\label{nst}
\eeqn
where $T^a$ are the generators of $SU(N)$, the covariant derivative is defined as 
\begin{center}
$\displaystyle 
\nabla_\mu=\partial_\mu-\frac{i}{2}A_\mu-iT^aA^a_\mu\,$,
\end{center}
$A_\mu$ and $A_\mu^a$ denote the $U(1)$ and $SU(N)$ gauge fields respectively, and the corresponding coupling constants are $g_1$ and $g_2$. The scalar fields $\varphi^{kA}$ have the color index 
$k=1,...,N$ and the flavor index $A=1,...,N$. Thus, $\varphi^{kA}$ can be viewed as an $N\times N$ matrix. The  $U(1)$ charges of $\varphi^{kA}$ are  $1/2$. 

\vspace{1mm}

Let us examine the potential of the theory (\ref{nst}) in more detail.
It consists of two non-negative terms and consequently the minimum of the potential is reached when 
both terms vanish. The last term proportional to $g_1^2\,$ forces $\varphi^A$ to develop a vacuum expectation value. 
One can choose $\varphi^{kA}$ to be proportional to the unit matrix, namely, 
\beq
\varphi_{vac}=\sqrt{\xi}\mbox{ diag } (1,1,...,1),
\eeq
where we use $N\times N$ matrix notation for $\varphi^{kA}$.
Then the last but one term vanishes automatically.

The above vacuum field spontaneously breaks both the gauge and flavor $SU(N)$ groups. However, 
it is invariant under the action of combined color-flavor global $SU(N)_{C+F}$. 
Therefore,  symmetry breaking pattern is 
\begin{center}
$\displaystyle U(N)_{\mbox{gauge}}\times SU(N)_{\mbox{flavor}}\rightarrow SU(N)_{C+F}$\,.
\end{center}
This setup was suggested in \cite{BH} and became known later as the color-flavor locking. 

The topological stability of non-Abelian strings in this model is due 
to the fact that $\pi_1(SU(N)\times U(1)/Z_N)\neq 0$. One combines the $Z_N$ center of $SU(N)$ with 
elements $e^{2\pi i k/N}$ of $U(1)$ to get windings in both groups simultaneously. 

The string solution \cite{GSY1} breaks the global symmetry of the vacuum as follows:
\beq
SU(N)_{C+F}\rightarrow SU(N-1)\times U(1)\, .
\label{sb}
\eeq  

As a result the orientational zero modes appear,  making the vortex non-Abelian. 
As is clear from the symmetry breaking pattern of Eq. (\ref{sb}) the orientational moduli belong to the quotient
\beq
\frac{SU(N)}{SU(N-1)\times U(1)}=CP(N-1)\, .
\eeq
Thus, the low-energy effective theory on the string world-sheet is described by the $CP(N-1)$ model. The action of the model was derived in \cite{GSY1};  it can be written as
\beq
S^{(1+1)}=\int d^2x\left[\frac{T_{\rm cl}}{2}(\partial_kz^i)^2+ r\,|\nabla_k\, n^l|^2\right]\,,
\eeq
where 
\beq
T_{\rm cl}=2\pi\xi
\label{ten}
\eeq
is the classical tension of the string, $z^i$ are two translational moduli in the perpendicular plane, 
$n^l$, $l=1,...,N$ are $N$ complex fields subject to the constraint 
\beq
|n^l|^2=1\,,
\label{n1}
\eeq
and $r$ is defined below.

The covariant derivative is
 \beq 
 \nabla_k=\partial_k-iA_k
 \label{covd}
\eeq
 and $k=(1,2)$ labels the world-sheet coordinates. 
The relation between two-dimensional coupling $r$ and a four dimensional coupling $g_2$ at 
the scale $\sqrt{\xi}$ is given by 
\beq
\displaystyle r=\frac{4\pi}{g_2^2}.
\label{gbeta}
\eeq
The field $A_k$ enters 
without kinetic term and is  auxiliary. It can be eliminated by virtue 
 of equations of motion and is introduced to make the $U(1)$ gauge invariance of the model explicit.

Let us count the number of degrees of freedom. The complex scalar fields give $2N$ real degrees of freedom, 
of which one is eliminated due to the constraint (\ref{n1}) and another one due to $U(1)$ gauge invariance. 
Thus, the total number of degrees of freedom is $2(N-1)$ which is precisely the number of degrees of freedom 
in the $CP(N-1)$ model.

To conclude this section we note that formation of non-Abelian strings leads to confinement of 
monopoles in the bulk theory. In fact, in the U$(N)$ gauge theories
strings are stable and cannot be broken. Therefore, confined monopoles are presented by junctions of 
two degenerate non-Abelian strings of different kinds,  see review \cite{SYrev} for details.
In the effective world-sheet theory on the string these confined monopoles are seen as CP$(N-1)$ kinks interpolating between distinct vacua.

\section{\boldmath{$CP(N-1)$} model at zero temperature}
\label{cpn1mazt}
\setcounter{equation}{0}

At large $N$ the model was solved \cite{W} in the $1/N$ approximation. Let us outline how this 
is done. 
The Lagrangian $\mathcal{L}$ of the $CP(N-1)$ model in the gauged formulation in the Euclidean space-time 
can be written as 
\beq
\mathcal{L}= |\nabla_k n^l|+\omega\left(|n^l|^2-r\right)\,,
\label{action}
\eeq
where we rescale the $n^l$ fields. In addition, we introduce a
parameter $\omega$ to enforce the constraint. Moreover, we replace the coupling $r$ with  the 't Hooft coupling constant $\lambda$, 
\beq
\lambda=\frac{N}{r}\,;
\eeq
$\lambda$ does not scale with $N$.

Since the $n^l$ fields appear quadratically in the action (\ref{action}) we can perform the Gaussian 
integration over them resulting in the equation for the effective potential $V$,
\beq
e^{-\hat{T} V}=\int d\omega\,dA_k\, \mbox{det}^{-N}\left(-(\partial_k-iA_k)^2+ \omega\right)
\mbox{exp} \left(\frac{N}{\lambda}\,\int d^2 x\,\omega\right)\,,
\label{det}
\eeq
where $\hat{T}$ stands for the (asymptotically infinite) Euclidean time.

Since integration over $\omega$ and $A_k$ cannot be done exactly we use a stationary phase approximation. Due to the Lorentz invariance we search for a point such that  $A_k=0$ and $\omega=\,$const. To find this stationary point we vary the Eq. (\ref{det}) with respect to $\omega$. The resulting equation is 
\beq
\lambda\int\frac{d^2k}{(2\pi)^2}\frac{1}{k^2+\omega}=1\,.
\eeq
Rewriting the bare coupling constant $\lambda$ in terms of the scale $\Lambda_{\rm CP}$ of the CP$(N-1)$ model
\beq
\frac{4\pi}{\lambda} = \ln{\frac{M_{\rm uv}^2}{\Lambda_{\rm CP}^2}}\,,
\label{Lambda}
\eeq
where $M_{\rm uv}$ is the ultra-violet cutoff,
we finally find that 
\beq
\omega = \Lambda_{\rm CP}^2 \,.
\label{lambda}
\eeq
Thus, the  vacuum value of $\omega$ does not vanish. Looking  at  Eq. (\ref{action}) one can see that 
a positive value of $\omega$ means that a mass for the fields $n^l$ is dynamically generated. 

\begin{figure}[h!]
\includegraphics[width=13.5cm]{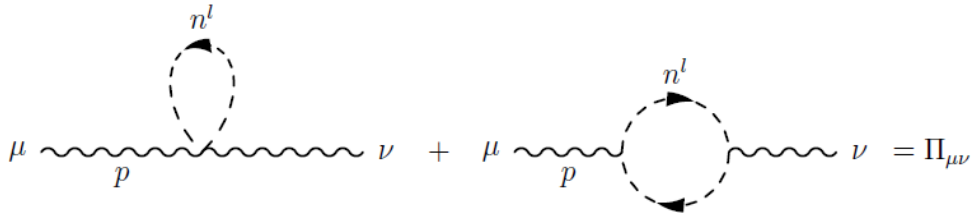}
\caption{\small Feynman diagrams contributing to kinetic term of photon field}
\label{diag1} 
\end{figure}

To determine the spectrum of the theory one has to expand the effective action Eq. (\ref{action}) around the saddle point and consider field fluctuations in the quadratic approximation. Linear terms vanish. Terms that are cubic and higher are suppressed by powers of $1/\sqrt{N}$.  
Two Feynman diagrams in Fig.~\ref{diag1} give rise to the  
kinetic term for the U(1) gauge field. 

Gauge invariance requires the answer to be 
\beq
\Pi_{\mu\nu}=\Pi(p^2)\left(p^2g_{\mu\nu}-p_\mu p_\nu\right)\,.
\label{kinetic}
\eeq
The meaning of Eq. (\ref{kinetic}) is simple. It represents the kinetic energy of 
the gauge field written in momentum space. Thus, what was introduced as an 
auxiliary field becomes a  propagating field. Calculation in Appendix B
reproduces Witten's result \cite{W}, $\Pi(0)=N/12\pi\Lambda_{\rm CP}^2$\,, which is interpreted as the inverse of the $U(1)$ charge squared of the $n^l$ fields. 

Massless photon in two dimensions produces the Coulomb potential 
between two charges at separation $R$,
\beq
V(R)=\frac{12\pi\Lambda^2}{N}\,R\,,
\label{coulomb}
\eeq
leading to a linear confinement of the $\bar{n}n$ pairs. Thus, the spectrum of the theory contains $\bar{n}n$ ``mesons" rather than free $n$'s. 

\begin{figure}
\centering
\includegraphics[width=6cm]{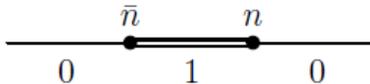}
\caption{\small Configuration of the string with two particles on it. Zero and one represent the true vacuum and the first quasivacuum respectively.}
\label{diag3} 
\end{figure}

It is instructive to present an alternative interpretation of this result. In \cite{W} it was shown that $n^l$ fields can be interpreted as kinks interpolating between different vacua. The vacuum structure of the $CP(N-1)$ model was studied in  \cite{W2}. According to this work the genuine vacuum is unique. There are, however, of the
order $N$ quasivacua, which become stable in the limit $N\rightarrow\infty\,,$ since the energy split between the neighboring quasivacua is $O(1/N)$. Thus, one can imagine the  $\bar{n}$ field interpolating between the true vacuum  and the first quasivacuum and the $n$ field returning to the true vacuum as in Fig. \ref{diag3}. The linear confining potential between the
kink and antikink is associated with the excess in the quasivacuum energy density compared to that in the genuine vacuum.

This two-dimensional confinement of kinks can be interpreted in terms of strings and monopoles of the bulk
theory, see \cite{GSY1}. The fine structure of the CP$(N-1)$ vacua  on the non-Abelian string means that
$N$ elementary strings are split by quantum effects and  have slightly different tensions.
Therefore, the monopoles, in addition to the four dimensional confinement, (which ensures that
they are attached to the string) acquire a two-dimensional confinement along the string. 
The monopole and antimonopole connected by a string with larger tension form a mesonic bound state.

Consider a monopole-antimonopole pair  interpolating between strings 0 and 1, see Fig.~\ref{diag3}. 
The energy of the excited part of the string (labeled as $1$) is proportional to the distance as 
in Eq. (\ref{coulomb}). When it exceeds the 
mass of  two monopoles (which is of order of $\Lambda_{\rm CP}$) then the second monopole-antimonopole pair 
appear breaking the excited part of the string. This gives an estimate for the typical 
length of the excited part
of the string, $R\sim N/\Lambda_{\rm CP}$.

 The above condition guarantees that there is enough energy 
in the ``wrong string" to produce a pair of kinks. 
However, the probability of this process, string breaking,  (which can be inferred from the false vacuum decay theory)
is proportional to $\exp(-N)$, i.e. dies off exponentially at large $N$.

\section{The Coulomb/confinement phase}
\label{seclargeL}
\setcounter{equation}{0}

In order to consider closed non-Abelian strings of length $L$ we compactify the space dimension;
 in other words, we study CP$(N-1)$ model (\ref{action})
on a strip of the finite length $L$ with periodic boundary conditions. 

In Euclidean formulation considering a model at finite length is equivalent to considering the 
model at finite temperature. The correspondence between the length of the string and the temperature is given by 
\beq
L=\beta\,,
\eeq
where $\beta$ is the inverse temperature. Thus, the limit of infinite length is the same as the 
limit of zero temperature. 

To solve the CP$(N-1)$ model on a finite strip we use large-$N$ approximation. 
The $CP(N-1)$ model at finite temperature in the large-$N$ approximation was solved previously by
Affleck \cite{Aff1}, see also \cite{Laz} and \cite{Actor} for reviews. 
Although we use a different regularization, our results match those obtained in \cite{Aff1}. There are two 
important differences, however. The first one is related to the interpretation of the photon mass. In \cite{Aff1}
the emergence of the photon mass is interpreted as a phase transition into the deconfinement phase 
already at $L=\infty$.
We give a different interpretation of the photon mass (see Sec.~\ref{secphotonmass}); we do not detect
any phase transition at  $L=\infty$. We interpret the large $L$ phase ($L > 1/ \Lambda_{\rm CP}$)
as a Coulomb/confinement phase, much in the same way as at infinite $L$ \cite{W}.

The second difference with Ref. \cite{Aff1} is that  we find a phase transition at $L \sim 1/ \Lambda_{\rm CP}$
into a deconfinement phase in the limit $N\to\infty$, see Sec.~\ref{secdeconfinement}. 
This is a weak coupling phase. 
In this phase the global SU$(N)$ is broken and the CP$(N-1)$ model does not develop a mass gap. The gauge
field remains auxiliary and no Coulomb/confining potential is generated. 

At large but {\em finite} $N$ we expect  
the phase transition to become a rapid crossover. The spontaneous breaking of the global SU$(N)$ symmetry
is in a contradiction with the Coleman theorem \cite{C},  stating that there can be no massless non-sterile
particles in $1+1$ dimensions. Therefore we expect that the ``would be Goldstone'' states
of the broken phase acquire small masses suppressed in the  large-$N$ limit.

To solve the CP$(N-1)$ model  we use the
mode expansion with the periodic boundary conditions. The open string setup involves the Dirichlet 
boundary conditions. For example, for open string the expansion (\ref{ANOexp}) is modified.
It acquires $L$-independent  terms coming from the energy associated with boundaries.  We  limit ourselves to a 
closed string in this paper.

\subsection{Large-\boldmath{$N$} solution}

Our starting point is Eq. (\ref{action}). 
Integrating out $n^l$ fields, one arrives at the same Eq. (\ref{det}) as in the infinite $L$ case. 
However, now we take into account the gauge holonomy around the compact dimension.
Following \cite{Aff1} we choose the gauge $$A_1=0$$
and look for minima of the potential with  $A_0=\,$ const and $\omega=\,$ const. The mode expansion in 
(\ref{det}) gives for the orientational part of the string energy in (\ref{NAexp})
\beq
E_{\rm orient}(L) = \frac{N}{2\pi}\, \sum_{k=-\infty}^{\infty}\int_{-\infty}^{\infty} dq_1
 \ln{\left\{q_1^2+\left(\frac{2\pi k}{L}+A_0\right)^2+\omega\right\}}\,.
\label{generalE}
\eeq

To calculate (\ref{generalE})  we follow \cite{Haw} and use the  zeta function regularization. 
Details of our calculation are presented in  Appendix A.
 Here we give the  final result for the string vacuum energy, 
\beq
E_{\rm orient}(L)=\frac{NL\omega}{4\pi}\left[1-\ln\frac{\omega}{\Lambda_{CP}^2}
-8\sum_{k=1}^{\infty} \frac{K_1(kL\sqrt{\omega})}{kL\sqrt{\omega}}\cos{kLA_0}\right]\,,
\label{Vl}
\eeq
where $K_1$ is the modified Bessel function of the second kind (also known as the Macdonald function). An important feature of this 
expression is the appearance of a non-trivial potential for the photon field. We will dwell on 
this issue in the next subsection.

To find the saddle point we extremize the expression (\ref{Vl}) with respect to $\omega$ and $A_0$, which results in the following equations:
\beqn
\frac{\d E_{\rm orient}}{\d A_0} 
&=&
\frac{2NL\sqrt{\omega}}{\pi}\sum_{k=1}^{\infty}K_1(Lk\sqrt{\omega})\sin{LkA_0}=0\, ,
\label{va}
\\[3mm]
\log {\frac{\omega}{\Lambda_{\rm CP}^2}}
&=&
4\sum_{k=1}^{\infty}K_0(Lk\sqrt{\omega})\cos{LkA_0}\, ,
\label{vm}
\eeqn
where the logarithmic term in the left-hand side of Eq. (\ref{vm}) is the renormalized inverse coupling $1/\lambda$. The logarithmic 
integral over momentum is regularized in the infrared by $\omega$.

\begin{figure}[h!]
\centering
\includegraphics[width=7.5cm]{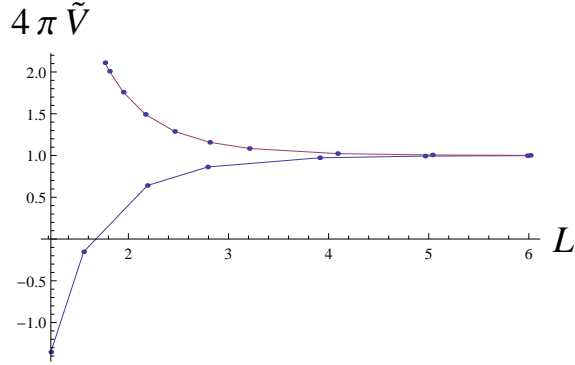}
\caption{\small Effective potential (in units of $\Lambda_{\rm CP}^2$) as a function of length.}
\label{v-b-g}
\end{figure}

Equation (\ref{va}) yields the solution of the form $LA_0=\pi\, l$, where $l\in\mathbb{Z}$. However,  from the Eq. (\ref{Vl}) it is clear that the solution with $LA_0=2\pi\, l$ lies lower in energy than the solution with $LA_0=(2l-1)\pi$ and is, thus, physical. We take $A_0=0$ as a solution of (\ref{va}).
Our result for the orientational string energy is shown in Fig. \ref{v-b-g}, where $\tilde{V}=E_{\rm orient}/L$.

\begin{figure}[h!]
\centering
\includegraphics[width=7.5cm]{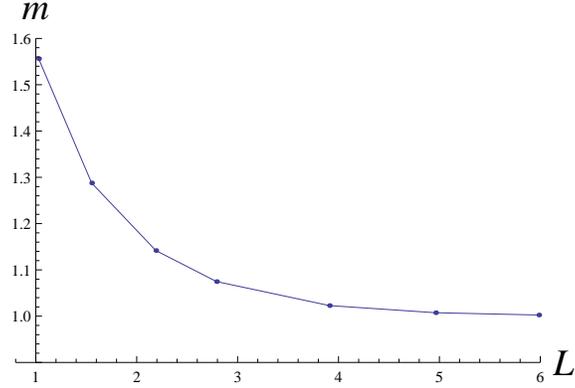}
\caption{\small Mass (in the units of $\Lambda$) of fields $n^l$ as a function of $L$.}
\label{ml}
\end{figure}

Equation (\ref{vm}) yields a nonvanishing value of $\omega$ which we interpret -- as in the case of zero temperature -- as mass generation for the $n^l$ fields. The dependence of the mass on the string length $L$  is shown in 
Fig. \ref{ml} where we put
\beq
 \sqrt{\omega} \equiv m\,.
 \eeq
  One can see that the $n^l$ field mass 
increases with decreasing $L$. 

In the limit  $L\gg 1/\Lambda_{\rm CP}$ the modified Bessel functions in (\ref{Vl}) 
exhibit exponential fall-off at  large $L$. To determine the leading non-trivial correction to
the string energy we can use the ``zeroth-order'' solution  $\omega\approx \Lambda^2_{\rm CP}$
of the equation (\ref{vm}) for the vacuum expectation value (VEV) of $\omega$. Clearly this ``zeroth-order'' solution coincides with the 
VEV of $\omega$ in the infinite volume, see (\ref{lambda}). 
 For the total string energy we obtain
\beq
E(L) = \left(2\pi \xi+ \frac{N}{4\pi}\,\Lambda_{\rm CP}^2\right)\,L  -\frac{\pi}{3}\,\frac1L -
 N\sqrt{\frac{2}{\pi}}\,\sqrt{\frac{\Lambda_{\rm CP}}{L}}\,e^{-\Lambda_{\rm CP}L}
+ \cdots .
\label{energylargeL}
\eeq
In Eq. (\ref{energylargeL})  we included the classical string tension $2\pi\xi L$, its renormalization due to
vacuum fluctuations in $CP(N-1)$ 
(i.e. $(N/4\pi)\,\Lambda_{\rm CP}^2\,L$),
  and 
the contribution of the translational modes which give the  standard  L\"uscher term.
This result was quoted in Sec. \ref{intro}, see Eq.~(\ref{flargeL}). 

We see that the quantum fluctuations
of the orientational moduli contribute both to the renormalization of the string tension (the linear in $L$ term
in (\ref{energylargeL})) and to the function $f(\Lambda_{CP}L)$ in (\ref{NAexp}). As was expected, in the theory with a mass gap the 
contribution of orientational moduli to the $L$-dependent part of the string energy is exponentially
suppressed at large $L$.

Let us note, that the case of an open non-Abelian string was previously considered  in \cite{Milekhin}.
The results of \cite{Milekhin} show the presence of long range $1/L$ effects coming from the orientational sector
even at large $L$ where the theory has a mass gap. We disagree with these results and believe that
orientational long range forces in the large-$L$ phase are spurious and are associated with the boundary energy
somehow induced \cite{Milekhin} by the Dirichlet boundary conditions  rather than with the string itself.

\subsection{The photon mass}
\label{secphotonmass}

The $A_0$-dependence in  the potential  (\ref{Vl})  ensures that the gauge field acquires a mass \cite{Aff1}.
It is quite natural to expect that the photon becomes massive at non-zero temperature. Physically this means the
Debye screening.

Expanding (\ref{Vl}) at large $L$ we can write down an effective action for the U(1) gauge field,
\beq
S_{\rm gauge} =\int d^2 x \left\{ \frac{1}{4e^2}\,F_{kl}^2 - 
 N\sqrt{\frac{2}{\pi}}\,\sqrt{\frac{\Lambda_{CP}}{L^3}}\,e^{-\Lambda_{\rm CP}L}\, \cos{A_0 L}
+\cdots \right\}.
\label{phaction}
\eeq

The kinetic term for the gauge field at non-zero temperature is calculated in Appendix B. To 
calculate the photon mass to the leading
order in $\exp{(-\Lambda_{\rm CP}L)}$ we need the expression for the gauge coupling $e^2$ in the limit
$L\to\infty$, namely,
\beq
\frac{1}{e^2}\approx \frac{N}{12\pi \Lambda_{\rm CP}^2},
\eeq
see Sec.~\ref{cpn1mazt}. Expanding (\ref{phaction}) to the quadratic order in $A_0$ we arrive at 
\beq
m_{A}^2\approx 12\Lambda_{\rm CP}^2\,\sqrt{2\pi\Lambda_{\rm CP} L}\;e^{-\Lambda_{\rm CP} L}\,.
\label{photonmass}
\eeq
for the photon mass.
Note, that the non-zero photon mass at finite temperature does not break gauge invariance  since Lorentz symmetry is explicitly broken, see \cite{Aff1}. 

The photon becoming massive  was the reason for the claim \cite{Aff1} that at non-zero temperature the CP$(N-1)$ model is
in the deconfinement phase. We give a different interpretation for this effect.  

We treat 
the quasivacua as the strings of different tension. Kinks and antikinks interpolate between true 
vacuum and the first quasivacuum. The Debye screening due to a finite photon mass now can be interpreted as a 
breaking of the confining string between kink and antikink in the thermal medium (through picking up a kink-antikink pair from the thermal bath). Note, that unlike pair-production from the vacuum, this process is not suppressed as $\exp (-N)$.

The kink-antikink potential has the form
\beq
V(R) =e^2\,R\,e^{-m_A R}\,,
\label{confscreenpot}
\eeq
where $R$ is the  kink-antikink separation. It is still linear at small $R$, while the exponential 
suppression at large $R$ can be understood  as a breaking of the confining string due to creation of 
a kink-antikink pair from the thermal bath.
Therefore, we still
interpret the large $L$ phase as a Coulomb/confinement phase.

A similar question can be addressed in QCD. Do we have confinement of quarks in QCD? We believe 
that the answer is positive. However, the confining string can be broken by quark-antiquark production.
We suggest a similar interpretation for the CP$(N-1)$ model at non-zero temperature. 

%


If $L$ is very large (very low temperatures) the thermal string breaking can 
be ignored, however once $L$ reduces below $\log{N}/\Lambda_{\rm CP} $ the thermal breaking
becomes operative.
%
%
\subsection{Small length limit}
\label{secsmallL}

As was already mentioned, we will show in the next section that once $L$ decreases  below $1/\Lambda_{CP}$
our CP$(N-1)$ model undergoes a phase transition into the deconfinement phase. To prove this we 
calculate the vacuum energy in the deconfinement phase in the next section and show that it lies
below that in the Coulomb/confinement phase.

In order to make this comparison we will examine Eqs. (\ref{Vl}) and (\ref{vm}) in the low-$L$ limit. 
These expressions determine the
vacuum energy and the $\omega$ expectation value in the Coulomb/confinement phase. 

Assuming that $L^2\omega \ll 1$ we can use the following approximation for the sum of the modified Bessel functions 
(see Eq. (8.526) in \cite{GR})
\beq
\sum_{n=1}^{\infty}K_0(ny)\approx\frac{\pi}{2y}+\frac{1}{2}\ln\frac{y}{4\pi}+\frac{\gamma}{2}+O(y^2)\,,
\label{k0}
\eeq
where $\gamma\approx0.577\,$ is the Euler-Mascheroni constant. Consequently, we get from (\ref{vm})
\beq
\ln{\frac{\sqrt{\omega}}{\Lambda_{CP}}} = 2\left[\frac{\pi}{2L\sqrt{\omega}}+\frac{1}{2}
\ln\frac{L\sqrt{\omega}}{4\pi}+\frac{\gamma}{2}\right]\,,
\eeq
or approximately
\beq
\ln{\frac1{\Lambda_{CP} L}} = \frac{\pi}{L\sqrt{\omega}}.
\label{omegaeq}
\eeq

Now the logarithmic integral which determines the renormalized inverse coupling $1/\lambda$ is regularized
 in the
infrared by $1/L$ rather than by $\sqrt{\omega}$ (which is the case in the large-$L$ limit). 
This gives us the $\omega$ expectation value,
\beq
\sqrt{\omega} = \frac{\pi}{L}\, \frac1{\ln{(1/\Lambda_{\rm CP} L})} +\cdots .
\label{omegasmallL}
\eeq
Equation (\ref{omegasmallL})  justifies our approximation  $L^2\omega \ll 1$ at  $L\ll 1/\Lambda_{CP}$.
Note also that at  $L\ll 1/\Lambda_{CP}$
the coupling constant is small -- it is frozen at the scale $1/L$ (the logarithm in the left-hand side of (\ref{omegaeq})
is large), so the theory is at weak coupling. 

To find the orientational energy in this limit we need to find an approximate expression 
for the sum of the modified Bessel functions that appears in  (\ref{Vl}), 
\beq
S_E=\frac{2L\sqrt{\omega}}{L\pi}\sum_{k=1}^{\infty} \frac{K_1(kL\sqrt{\omega})}{k}\,.
\eeq
Derivative of the modified Bessel functions satisfies the following relation 
(see Eq. (9.6.28) in \cite{Abr}):
\beq
K_1^{\prime}(x)=-K_0(x)-\frac{K_1(x)}{x}\,.
\eeq
Let us introduce a notation, 
\beq
S_1(x)=\sum_{k=1}^{\infty} \frac{K_1(kx)}{k}\,.
\eeq
Then 
\beq
(xS_1(x))^{\prime}=-x\sum_{k=1}^{\infty} K_0(kx)\stackrel{(\ref{k0})}{\approx}
-\frac{\pi}{2}-\frac{x}{2}\ln\frac{x}{4\pi}-\frac{x\gamma}{2}+O(x^3)\,.
\eeq
Integrating this expression one finds
\beq
xS_1(x)\approx-\frac{x\pi}{2}-\frac{x^2}{4}\ln\frac{x}{4\pi}-\frac{x^2}{8}(2\gamma-1)+\mbox{const}+O(x^4)
\eeq
The behavior of the modified Bessel function at small values of the argument is given by 
(see Eq. (9.6.9) in \cite{Abr}) 
\beq
K_1(x)\sim\frac{1}{x}\,.
\eeq
Thus, the sum $S_1(x)$ can be approximated as follows:
\beq
S_1(x)\approx \sum_{k=1}^{\infty} \frac{1}{xk^2}=\frac{\pi^2}{6x}\,.
\eeq
Hence the constant appears to be $\pi^2/6$. Now we are ready to present 
the approximate expression we  seek for,
\beq
S_E=\frac{2}{L\pi}L\sqrt{\omega}S_1(L\sqrt{\omega})\approx\frac{\pi}{3L}-\sqrt{\omega}
-\frac{L\omega}{2\pi}\ln\frac{L\sqrt{\omega}}{4\pi}-\frac{L\omega}{4\pi}(2\gamma-1)\,.
\eeq
With this approximation we arrive at the orientational energy 
\beq
E_{\rm orient}(L) = -\frac{\pi}{3}\,\frac{N}{L} + N\,\sqrt{\omega} - \frac{N}{2\pi}\, 
\omega L\,\ln{\frac{1}{\Lambda_{CP} L}} +\cdots
\label{EsmallL}
\eeq
Substituting here the VEV of $\omega$, see (\ref{omegasmallL}), we get
\beq
E_{\rm orient}(L) = -\frac{\pi}{3}\,\frac{N}{L} + \frac{\pi}{2}\,\frac{N}{L}\,
 \frac1{\ln{(1/\Lambda_{CP} L)}}
 +\cdots .
\label{energysmallL}
\eeq

The first term here is the L\"uscher term proportional to the number of orientational degrees
of freedom $2(N-1)\approx 2N$ (in the large $N$ limit). It gets corrected by an infinite
series of powers of inverse logarithms $\ln{(1/\Lambda_{CP} L)}$, if we naively extend the
 Coulomb/confinement
phase into the region of small $L$.  We will show in the next section that
in fact the theory undergoes  a phase transition into a different phase, with a lower energy.

\section{Deconfinement phase}
\label{secdeconfinement}
\setcounter{equation}{0}

Classically CP$(N-1)$ model has $2(N-1)$ massless states which can be viewed as Goldstone states
of the broken SU$(N)$ symmetry. Indeed, classically the vector $n^l$ satisfies a  fixed length
condition, $|n|^2 =r$, see (\ref{action}). Thus classically $n^l$ acquires a VEV breaking SU$(N)$ symmetry.

However, as was shown above, in the strong coupling  large $L$ domain the spontaneous symmetry breaking
does not occur, in much the same way as in the infinite-$L$ limit, see \cite{W}. At strong coupling the
vector $n^l$ is smeared all over the vacuum manifold due to strong quantum fluctuations. The theory has a mass gap, moreover the number of the 
massive $n$-fields  becomes  $2N$. Effectively the classical constraint $|n|^2 =r$ is lifted,
see \cite{W}.

At small $L$ the theory enters a weak coupling regime so we expect occurrence of the classical
 picture in the limit $N\to\infty$. To study this possibility we assume that one component of the field
$n^l$, say $n_0\equiv n$ can develop a VEV. Then we integrate over all other components 
of $n^l$ ($l$=1,2,...) keeping 
the  fields  $n$ and $\omega$ as a background. Note, that a similar method was used in \cite{GSYphtr}
for studying  phase transitions in the CP$(N-1)$ model with twisted masses.

Now, instead of (\ref{EsmallL}), we get
\beq
E_{\rm orient}(L) = \omega L\,|n|^2-\frac{\pi}{3}\,\frac{N}{L} - \frac{N}{2\pi}\, 
\omega L\,\ln{\frac{1}{\Lambda_{\rm CP} L}} +\cdots ,
\label{EsmallLh}
\eeq
where the ellipses stand for higher terms in $L^2\omega$. Note, that here we drop the contribution associated 
with the integration over the constant $n$ (the second term in (\ref{EsmallL})) because we introduce 
$n_0$ as a constant background field (in other words, we drop the term with $k=0$ in (\ref{generalE})).

Minimizing over $\omega$ and $n$ we arrive at the equations
\beqn
|n|^2&=&\frac{N}{2\pi}\,\ln{\frac{1}{\Lambda_{CP} L}} +\dots ,
\\[2mm] 
\nonumber
\omega\, n&=&0\,.
\label{csp}
\eeqn

The solution to these equations with nonzero $n_0$ read
\beq
|n|^2=\frac{N}{2\pi}\,\ln{\frac{1}{\Lambda_{CP} L}}, \qquad \omega=0\,.
\label{decphasesol}
\eeq
We see that the mass gap $\omega$ is not generated. Substituting this in (\ref{EsmallLh}) we get that
the orientational energy reduces just to  the  L\"uscher term, namely
\beq
E_{\rm orient}(L) = -\frac{\pi}{3}\,\frac{N}{L} .
\label{EsmallLLush}
\eeq

This energy is lower than the one in (\ref{energysmallL}). Therefore,  we conclude that at $L\sim 1/\Lambda_{CP}$ the theory undergoes  a phase transition into the phase with the broken SU$(N)$ symmetry.
This ensures the presence of $2(N-1)$ Goldstone states $n^l$, $l=1,...(N-1)$. The photon remains an
auxiliary field, no kinetic term is generated for it. As a result, there is no Coulomb/confining
linear rising potential between the $n$-states. The phase with the broken SU$(N)$ is a deconfinemet phase.
Since $|n^l|$ is positively defined Eq. (\ref{decphasesol}) shows that this phase appears at 
$L< 1/\Lambda_{\rm CP}$.

\begin{figure}[h!]
\centering
\includegraphics[width=10cm]{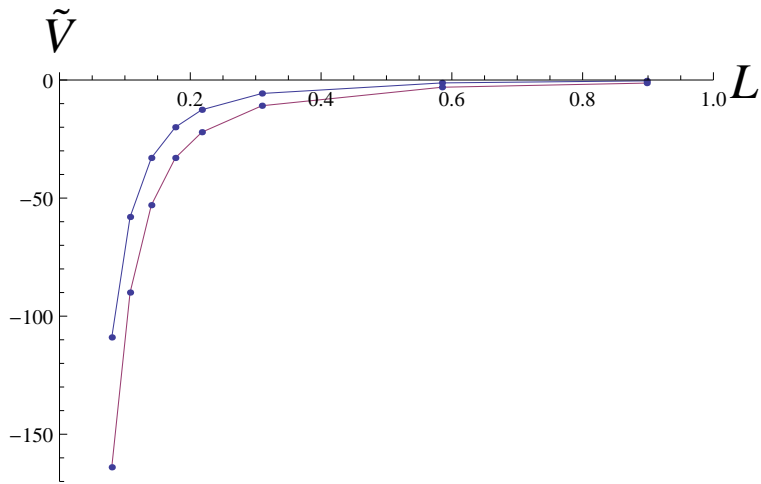}
\caption{\small Comparison of orientational energies in both phases. The L\"uscher term always lies lower. We set $\Lambda_{\rm CP}=1$.}
\label{v-vs-v0} 
\end{figure}

The results of numerical calculations are in agreement with our conclusions. The 
relation between orientational energies in both phases is shown in Fig. (\ref{v-vs-v0}).
One can see that the L\"uscher term energy is lower and is thus physical.

The phase with the broken symmetry in two dimensions can occur only in the limit $N\to\infty$. As was already 
explained, if $N$ is large but finite this would contradict the Coleman theorem \cite{C}. Therefore, 
we expect that at large but finite $N$ the phase transition becomes a rapid crossover.
In particular, we expect that the $n^l$  fields  are not strictly massless. They 
have small masses suppressed by $1/N$. 

To conclude this section let us note that the $CP(N-1)$ model compactified on a cylinder with the so-called 
twisted boundary conditions was studied in \cite{Unsal}. No phase transition was found; moreover,
it was shown that the theory has a mass gap which shows no $L$-dependence and is determined entirely  
 by $\Lambda_{\rm CP}$. We believe that our results are not in contradiction with those obtained 
in \cite{Unsal}, because at finite $L$ the boundary conditions matter: they can be crucial. In particular, the 
twisted boundary conditions
can be viewed as a gauging of the global SU$(N)$ group with a constant gauge potential. Then the global SU$(N)$ is
explicitly broken. This model should be considered as distinct  as compared to the CP$(N-1)$
model with the periodic boundary conditions studied in this paper.

\section{Supersymmetric CP\boldmath{$(N-1)$} model with no compactification}
\setcounter{equation}{0}
\label{susyrev}

Non-Abelian strings were first found in \ntwo supersymmetric QCD with the U$(N)$ gauge group and 
$N_f=N$  quark hypermultiplets  \cite{HT1,ABEKY,SYmon,HT2}, see \cite{Trev,Jrev,SYrev,Trev2}
for reviews. In much the same way as for non-supersymmetric case the internal dynamics of orientational
zero modes of non-Abelian string is described by two-dimensional CP$(N-1)$ model
living on the string world-sheet. The string solution is 1/2-BPS saturated; therefore the two-dimensional model under consideration is
\ntwot supersymmetric. In this section we briefly review the  large-$N$ solution of \ntwot \cpn model
in infinite space \cite{W}. In the next section we will present the large-$N$ solution of the model on a strip of
a finite length $L$ (cylindrical compactification).

The bosoinc part of the 
action of the \cpn model is given by
\beqn
S_{bos}&=&\int d^2x\Big[ |\nabla_in^l|^2+\frac{1}{4e^2}F_{ij}^2+\frac{1}{e^2}|\partial_i\sigma|^2+\frac{1}{2e^2}D^2 \nonumber \\[3mm]
&+&2|\sigma|^2|n^l|^2+iD(|n^l|^2-r_0)\Big]\,,
\eeqn
where the covariant derivative is defined as 
$\nabla_i=\partial_i-iA_i$ and $\sigma$ is a complex scalar field,
the scalar superpartner of $A_i$. Moreover, $r_0$ is the bare coupling constant. In the limit 
$e^2\rightarrow\infty$ the gauge field $A_i$ and  $\sigma$ become auxiliary fields. 
$D$ stands for the $D$ component of the gauge multiplet. The factor   $i$ is due to the passage to the
Euclidean notation.

The fermionic part of the action takes the form 
\beqn
S_{ferm}&=&\int d^2x \Big[\bar{\xi}_{lR}i(\nabla_0-i\nabla_3)\xi^l_R+\bar{\xi}_{lL}i(\nabla_0+i\nabla_3)\xi^l_L\nonumber \\[3mm]
&+&\frac{1}{e^2}\bar{\lambda}_Ri(\nabla_0-i\nabla_3)\lambda_R+\frac{1}{e^2}\bar{\lambda}_Li(\nabla_0+ i\nabla_3)\lambda_L\nonumber \\[3mm]
&+&\left(i\sqrt{2}\sigma\bar{\xi}_{lR}\xi^l_L+i\sqrt{2}\bar{n}_l(\lambda_R\xi^l_L-\lambda_L\xi^l_R)+\mbox{H.c.}\right)\Big]\,,
\label{susyfermac}
\eeqn
where the fields $\xi^l_{L,R}$ are the fermion superpartners of $n^l$ and $\lambda_{L,R}$ belong to the gauge multiplet. In the limit $e^2\rightarrow\infty$ they enforce the following constraints: 
\beq
\bar{n}^l\xi^l_L=0\,, \quad \bar{n}^l\xi^l_R=0\,.
\eeq
The field $\sigma$ is auxiliary and can be eliminated, namely,
\beq
\sigma=-\frac{i}{\sqrt{2}r_0}\bar{\xi}_{lL}\xi^l_R\,.
\eeq

\subsection{Large-\boldmath{$N$} solution}

The \ntwot supersymmetric  \cpn model  was solved in the large-$N$ limit by Witten \cite{W}, 
see also \cite{SY2}. In this section we briefly 
review this solution. 

Since both fields $n^l$ and $\xi^l$ appear quadratically we can integrate them out. This produces two determinants,
\beq
\mbox{det}^{-N}\left(-\partial_i^2+iD+2|\sigma|^2\right)\mbox{det}^N\left(-\partial_i^2+2|\sigma|^2\right)
\label{sdetzero}
\eeq
The first determinant comes from the boson $n^l$ fields, while the second comes from the fermion $\xi^l$ fields. 
Note 
that if $D=0$ the two contributions obviously cancel each other, and supersymmetry is 
unbroken. As before, the non-zero values of $iD+2|\sigma|^2$ and $2|\sigma|^2$ can be interpreted as non-zero values of the mass of $n^l$ and $\xi^l$ fields,
and we put $A_k=0$.

 The final expression for the effective potential
is given by (see, for example, \cite{SY2})
\beq
V_{\rm eff}=\int d^2x\frac{N}{4\pi}\left[-(iD+2|\sigma|^2)\ln\frac{iD+2|\sigma|^2}{\Lambda_{CP}^2}
+iD+2|\sigma|^2\ln\frac{2|\sigma|^2}{\Lambda_{CP}^2}\right],
\eeq
where the logarithmic ultraviolet divergence of the coupling constant is traded for the scale $\Lambda_{\rm CP}$.

To find a saddle point we minimize the potential with respect to $D$ and $\sigma$, which yields the following set of equations:
\beqn
\ln\frac{iD+2|\sigma|^2}{\Lambda_{CP}^2}=0\,,\nonumber \\[3mm]
\ln\frac{iD+2|\sigma|^2}{2|\sigma|^2}=0\,,
\label{infLeqs}
\eeqn
The solution to these equations is
\beq
D=0
\label{D=0infL},
\eeq
which shows that supersymmetry is not broken. The VEV of $\sigma$ is    
\beq
\sqrt{2}\sigma = \Lambda_{CP} \,e^{\frac{2\pi k}{N}i}, \qquad  k=0,..., (N-1).
\label{sigma0}
\eeq
We see that  $\sigma$ develops a VEV giving masses to the $n^l$ fields and their fermion superpartners
$\xi^l$. The phase factor in the right-hand side of (\ref{sigma0}) does not follow from
(\ref{infLeqs}). It comes from the  broken chiral U(1) symmetry. The axial anomaly breaks it
down to   $Z_{2N}$. The field  $\sigma$ has the chiral charge 2. This explains the phase factor in 
(\ref{sigma0}). Once $|\sigma|$ has a nonzero VEV the anomalous symmetry breaking ensures
that the theory has $N$ vacuum states. Clearly this fine structure cannot be seen in the large $N$
approximation since the phase factor is a $1/N$ effect.

In full accord with the Witten index,  the solution above has $N$ vacua, each with the vanishing energy.

\begin{figure}[h!]
\centering
\includegraphics[width=13.5cm]{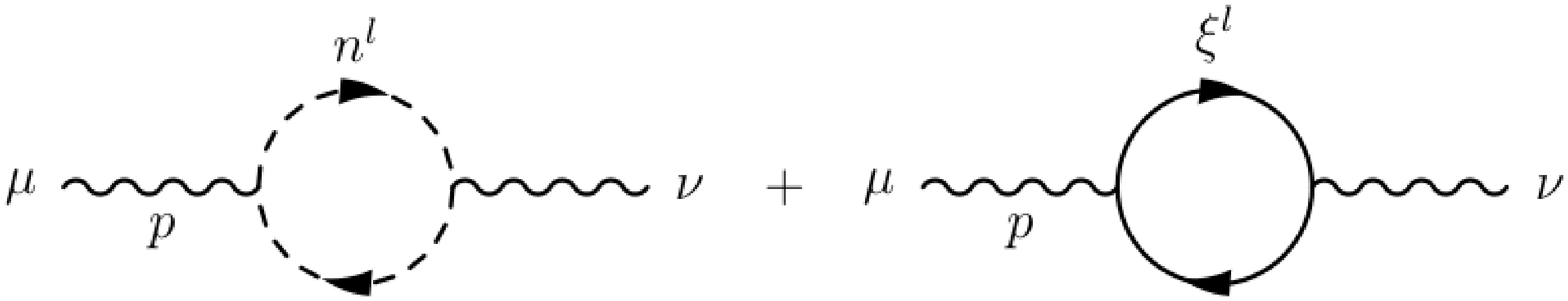}
\caption{Feynman diagrams contributing to the kinetic term of the photon}
\label{fds1} 
\end{figure}

Consider now the vector multiplet.
 In much the same way as in the non-supersymmetric case, photon becomes a 
propagating field. To find the renormalized gauge coupling one needs to 
evaluate two Feynman diagrams shown in the Fig.{\ref{fds1}}.
Details of the appropriate calculation are given in Appendix C. The result is 
\beq
\frac{1}{e^2}=\frac{N}{4\pi}\frac{1}{\Lambda_{\rm CP}^2}\,.
\eeq 

Through the coupling to the $\mbox{Im}\,\sigma$ 
(due to the chiral anomaly) now the photon acquires a mass. Moreover, the fermion fields 
$\lambda_{L,R}$ also become propagating, with the same mass as that of the photon, as 
required by supersymmetry.
The masses of the fields of the vector multiplet  are as follows \cite{W,SY2}:
\beq
m_{ph}=m_{\lambda_{L,R}}=m_{\mathrm{Re}\,\sigma}=m_{\mathrm{Im}\,\sigma}=2\Lambda_{CP}\,.
\eeq 

Since the photon became massive there is no linear rising Coulomb potential between the charged states.
There is no confinement in supersymmetric CP$(N-1)$ model even in the infinite volume limit. 
It has $N$ degenerate vacua which are interpreted as 
$N$ degenerate elementary non-Abelian strings in the four-dimensional bulk theory. 
In contrast to the non-supersymmetric case, the confined monopoles of the 
bulk theory, which are seen as kinks interpolating between the CP$(N-1)$ vacua, are free to move along the string,
see \cite{SYrev} for further details.

\section{Supersymmetric CP\boldmath{$(N-1)$} on a cylinder}
\setcounter{equation}{0}
\label{susy}

Now we compactify one space dimension and impose periodic boundary conditions, both for bosons and fermions, in order to preserve 
\ntwot super\-symmetry. We stress that this compactification can{\em{not}} be considered as thermal.
Non-zero temperature requires anti-periodic boundary conditions for fermions, which would
break supersymmetry explicitly.

The large-$N$ method in the case of \ntwot \cpn model works  similar to that in the non-supersymmetric case.
We compactify now the spatial coordinate $x_1$ and start from a slightly modified expression for the determinants in Eq. (\ref{sdetzero}). 
Choosing the $A_0=0$ gauge and assuming that $A_1$ is non-zero we write 
\beq
\mbox{det}^{-N}\left(-\partial_0^2-(\partial_1- iA_1)^2
+m_b^2\right)\mbox{det}^N\left(-\partial_0^2-(\partial_1- iA_1)^2+m_f^2\right)\,,
\label{sdetfin}
\eeq
where we introduced the following notation:
\beq
 m_b^2=iD+2|\sigma|^2, \qquad m_f^2=2|\sigma|^2.
\label{mbf}
\eeq 

 The 
evaluation of each of the determinants is no different from that in the non-supersymmetric case. Again we use 
the zeta-function method. Using expressions in Appendix C we can derive the effective potential,
\beqn
E&=&\frac{LN}{4\pi}\Big[-(iD+2|\sigma|^2)\ln\frac{iD+2|\sigma|^2}{\Lambda_{\rm CP}^2}+iD
+2|\sigma|^2\ln\frac{2|\sigma|^2}{\Lambda_{\rm CP}^2}\nonumber\\[2mm]
&-&8m_b^2\sum_{k=1}^\infty\frac{K_1(Lm_bk)}{Lm_bk}\cos{(LA_1k)}\nonumber\\[2mm]
&+&8m_f^2\sum_{k=1}^\infty\frac{K_1(Lm_fk)}{Lm_fk}\cos{(LA_1k)}\Big]\,,
\label{Esusymac}
\eeqn
Here the first line is just the effective potential at $L=\infty$, while the second and third 
lines are the finite-$L$ corrections due to bosons and fermions, respectively.

To find a stationary point we vary the above expression with respect to $A_1$, $D$ and $\sigma$.
The resulting equations are as follows:
\beqn
&&m_b\sum_{k=1}^\infty K_1(Lm_bk)\sin{(LA_1k)}-m_f\sum_{k=1}^\infty K_1(Lm_fk)\sin{(LA_1k)}=0\,,
\nonumber\\[2mm]
&&2\sigma\left[-\ln\frac{m_b^2}{m_f^2}+4\sum_{k=1}^\infty K_0(Lm_bk)\cos{(LA_1k)}
-4\sum_{k=1}^\infty K_0(Lm_fk)\cos{(LA_1k)}\right]=0\,,\nonumber\\[2mm]
&&-\ln\frac{m^2_b}{\Lambda_{CP}^2}+4\sum_{k=1}^\infty K_0(Lm_bk)\cos{(LA_1k)}=0\,.
\label{ssp}
\eeqn

Calculation of the gauge coupling constant at finite $L$ is also modified (see Appendix C).
As a result, we arrive at
\beq
\frac{1}{Ne^2}=\frac{1}{4\pi m_b^2}+\frac{L}{2\pi m_b}\sum_{k=1}^\infty K_1(Lm_bk)k\,,
\label{eLsusy}
\eeq
which reduces to $1/4\pi\Lambda_{\rm CP}^2$ in the limit $L\to\infty$.

Consider now the large $L$ limit, $L\gg 1/\Lambda_{\rm CP}$. Assuming that $m_b\sim m_f \sim \Lambda_{\rm CP}$
 (we confirm this below) we expand the string energy (\ref{Esusymac}) keeping the first exponentially small
term
\beqn
E&=&\frac{LN}{4\pi}\left\{-m_b^2\ln\frac{m_f^2}{\Lambda_{\rm CP}^2}+iD
+m_f^2\ln\frac{m_f^2}{\Lambda_{\rm CP}^2}
\right\}
\nonumber\\[3mm]
&-& N\,\sqrt{\frac{2}{\pi}}\,\left[\sqrt{\frac{m_b}{L}}\, e^{- m_b L}
- \sqrt{\frac{m_f}{L}}\, e^{- m_f L}\right]\,\cos{A_1 L} +\cdots.
\label{Esusy}
\eeqn
Taking derivatives with respect to $D$, $\sqrt{2}\bar{\sigma}$ and  $A_1$ we obtain
\beqn
&&
 -\frac{N}{4\pi} \log{\frac{m_b^2}{\Lambda_{CP}^2}^2}
+N\,\frac{1}{\sqrt{2\pi}}\,\frac{\exp{\left(- m_b L\right)}}
{\sqrt{ m_b L}}\,\cos{A_1 L} +\cdots =0,
\nonumber\\[3mm]
&&
\sqrt{2}\sigma\,\left\{
\frac{N}{4\pi} \log{\frac{m_f^2}{m_b^2}}
+N\,\frac{1}{\sqrt{2\pi}}\,\left[\frac{\exp{\left(- m_b L\right)}}
{\sqrt{ m_b L}} -\frac{\exp{\left(- m_f L\right)}}
{\sqrt{ m_f L}}\right]\,\cos{A_1 L} +\cdots \right\} =0,
\nonumber\\[3mm]
&&
\left\{\frac{\exp{\left(- m_b L\right)}}
{\sqrt{ m_b L}}-\frac{\exp{\left(- m_f L\right)}}{\sqrt{ m_f L}}\right\}\,\sin{A_1 L} +\cdots =0\,,
\label{susyeqs}
\eeqn
where the ellipses denote next-to-leading corrections in $1/Lm_b$ and  $1/Lm_f$.

The solution of these equations is as follows. The second and third equations are satisfied at
\beq
D=0
\label{D=0},
\eeq
which shows that supersymmetry is not broken. $A_1$ remains undetermined. 

With $D=0$ the first equation determines the $\sigma$ expectation value, namely,
\beq
\frac{N}{4\pi} \log{\frac{ 2|\sigma|^2}{\Lambda_{\rm CP}^2}} =
N\,\frac{1}{\sqrt{2\pi}}\,\frac{\exp{\left(-  \sqrt{2}|\sigma| L\right)}}
{\sqrt{  \sqrt{2}|\sigma| L}}\,\cos{A_1 L} +\cdots.
\label{sigmaeq}
\eeq
This equation seems to present a puzzle. It shows that the VEV of $\sigma$  depends on the parameter $A_1$, which is arbitrary.
If this were the case the theory would have a branch of vacua parametrized by the Polyakov line
\beq
 e^{\int dx_1 A_1} =e^{iA_1L},
\eeq
  which measures the holonomy
around the compact dimension. More exactly, the theory
would have $N$ branches of vacua, because $Z_{2N}$ symmetry ensures that the overall phase of $\sigma$ takes $N$
values $2\pi k/N$, $k=0,..., (N-1)$. This would contradict the Witten index argument which 
ensures that the number of vacua  is equal to $N$ for \ntwot supersymmetric CP$(N-1)$ model.

The resolution of this puzzle is that we should quantize the phase variable $A_1L$ (note that $\int dx_1 A_1$ depends only on time) as 
a function of the non-compact time. In the emerging quantum mechanics
the phase $A_1L$ is not fixed; instead, it is smeared all over the circle (in the ground state).
As a result, the $\cos{(A_1 L)}$ in (\ref{sigmaeq}) is averaged to zero and the $\sigma$ VEVs are given by 
\beq
\sqrt{2}\sigma = \Lambda_{\rm CP} \,e^{\frac{2\pi k}{N}i}, \qquad  k=0,..., (N-1).
\label{sigma}
\eeq
This  is exactly the same result as for $L=\infty$. All cosine functions
of $A_1L$ in the last equation in (\ref{ssp}) are averaged to zero, therefore the result in (\ref{sigma}) is exact and
does not depend on $L$. 

This result also can be understood by studying the exact twisted
superpotential of \ntwot CP$(N-1)$ model. In the infinite volume
it is given by \cite{AdDVecSal,W93,ChVa}
\beq
W(\sigma) = \frac{N}{4\pi}\left\{\sqrt{2}\sigma\, \log{\frac{ \sqrt{2}\sigma}{\Lambda_{\rm CP}}} -\sqrt{2}\sigma\right\}.
\label{sup}
\eeq
This superpotential has correct transformation properties with respect to the chiral U(1) symmetry. Namely, integrated over half of the superspace it is invariant under chiral symmetry up to a term which precisely reproduces the chiral anomaly. Now at finite length this superpotential in principle could have corrections 
proportional to powers of 
\beq
\exp{\left(-  \sqrt{2}\sigma L\right)}.
\eeq
However these corrections would spoil the transformation properties of the superpotential with respect to the chiral symmetry.
Therefore they are forbidden. As a result at finite $L$ the exact
superpotential of the theory is still given by (\ref{sup}). Critical points of this superpotential are given by (\ref{sigma}) and do not depend on $L$.
This matches our result obtained from large-$N$ approximation.

In particular, at small $L$ the theory is at weak coupling and can
be studied in the quasiclassical approximation. As we already mentioned 
CP$(N-1)$ model compactified on a cylinder with twisted boundary
conditions was studied in \cite{Unsal}. It is shown in \cite{Unsal}
that the mass gap at weak coupling is produced by fractional instantons
and does not depend on $L$
both in supersymmetric and non-supersymmetric cases. For our case
(periodic boundary conditions) the mass gap shows $L$-dependence
in non-supersymmetric case, while in the supersymmetric case it is
$L$-independent. The quasiclassical origin of this  behavior needs to be understood in the weak coupling domain of small $L$. This is left to 
a future work.

\vspace{3mm}

To conclude, in \ntwot supersymmetric CP$(N-1)$ model we have a {\em single} phase with the 
unbroken supersymmetry and
$N$ vacua.  Each vacuum has vanishing energy and parametrized by the VEV of $\sigma$ in Eq. (\ref{sigma}).  Unlike non-supersymmetric problem, this VEV is independent of $L$.

\section{The photon mass}
\setcounter{equation}{0}
\label{phma}

In this section we outline the photon mass calculation. 

The effective action for the gauge field can be written as \cite{SY2}
\beq
S_{\rm gauge} =\int d^2 x \left\{ \frac{1}{4e^2}\,F_{kl}^2 -\frac{N}{4\pi}\;
\log{\frac{\sigma}{\bar{\sigma}}}\;F^{*} \right\}\,,
\eeq
where the photon mixing with $\sigma$ is due to the chiral anomaly and 
\beq
F^*=\frac{1}{2}\epsilon_{ij}F^{ij}
\eeq
is the dual gauge field strength. In the case of infinitely long string the 
the gauge coupling and the photon mass were found \cite{SY2},
\beq
\frac{1}{e^2}=\frac{N}{4\pi}\frac{1}{\Lambda_{CP}^2}\,,
\eeq 
and 
\beq
m_{ph}=2\Lambda_{CP}\,,
\eeq 
respectively.
In Sec. \ref{susy} we derived the expression for the gauge coupling in the case of finite 
length, see  (\ref{eLsusy}). The corresponding expression for the photon mass  in the 
limit of $\Lambda_{CP} L\gg 1$ is  
\beq
m_{\rm ph}^2\approx(2\Lambda_{\rm CP})^2\left(1-\sqrt{2\pi\Lambda_{\rm CP} L}\;e^{-\Lambda_{\rm CP} L}\right)
\eeq
where 
we used the asymptotic expansion of the modified Bessel functions 
(see Eq. (9.7.2) in \cite{Abr}), 
\beq
K_1(x)\sim\sqrt{\frac{\pi}{2x}}e^{-x}\,.
\eeq
Since $K_0^{\prime}(x)=-K_1(x)$ we can also determine  the photon 
mass in the opposite limit of $\Lambda_{\rm CP} L\ll 1$,
\beqn
\sum_{k=1}^\infty K_1(kx)k&=&-\left(\sum_{k=1}^\infty K_0(kx)\right)^{\prime}\approx
\frac{\pi}{2x^2}-\frac{1}{2x}\,,\nonumber\\ [3mm]
m_{\rm ph}^2&\approx&\frac{\Lambda_{\rm CP} L}{\pi}\;(2\Lambda_{\rm CP})^2 \ll (2\Lambda_{\rm CP})^2 \,.
\eeqn

\section{Conclusions} 
\label{conclusions}

We studied two-dimensional CP$(N-1)$ model (both nonsupersymmetric and ${\mathcal N}= (2,2)$) compactified 
on a cylinder with circumference $L$ (periodic boundary conditions). We found the large-$N$ solution
for any value of $L$ and discussed in detail the large-$L$ and small-$L$ limits.

A drastic difference is detected in passing from the nonsupersymmetric to 
${\mathcal N}= (2,2)$ supersymmetric case.
In the former case in the large-$N$ limit we observe a phase transition at $L\sim \Lambda_{\rm CP}^{-1}$ (which is expected to become a rapid crossover at finite $N$).
At large $L$ the CP$(N-1)$ model develops a mass gap and 
is in the Coulomb/confinement 
phase, with exponentially suppressed finite-$L$ effects. At small $L$ it is in the deconfinement
phase; the orientational modes contribute to the L\"usher term. The latter becomes dependent on the rank of 
the bulk gauge group. 

In the supersymmetric CP$(N-1)$ model we have a different picture.  Our large-$N$ 
solution exhibits a single phase independently of the value 
of $L\Lambda_{\rm CP}$. For any value of this parameter a mass gap develops and supersymmetry
remains unbroken. So does the $SU(N)$ symmetry of the target space (i.e. it is restored). 
The mass gap turns out to be independent of the string length. The L\"uscher term is absent due to supersymmetry.

\section*{Acknowledgments}
\addcontentsline{toc}{section}{Acknowledgments}

\renewcommand{\theequation}{A.\arabic{equation}}
\setcounter{equation}{0}
 
 \renewcommand{\thesubsection}{A.\arabic{subsection}}
\setcounter{subsection}{0}

We are grateful to A. Vainshtein for illuminating insights. A.Y. is also grateful to M. Anber, 
T. Sulejmanpasic and M. \"Unsal for very useful discussions.

This work  is supported in part by DOE grant DE-SC0011842. 
The work of A.Y. was  supported by William I. Fine Theoretical Physics Institute  of the  University of 
Minnesota,
by Russian Foundation for Basic Research under Grant No. 13-02-00042a and by Russian State Grant for
Scientific Schools RSGSS-657512010.2. The work of A.Y. was supported by Russian Scientific Foundation 
under Grant No. 14-22-00281.

\section*{Appendix A: \\
Calculation of Zeta function}
\addcontentsline{toc}{section}{Appendices}

We define the zeta function of an operator $\Omega$ as follows: 
\beq
\zeta(s)=Tr\; \Omega^{-s}\,.
\eeq
The operator of interest is given in Eq. (\ref{det}),
\beq
\Omega=-(\partial_k-iA_k)^2+m^2\,,
\eeq
where instead of $\omega$ we write $m^2$. In the $A_1=0$ gauge the expression for the
zeta function takes the form
\beq
\zeta(s)=\frac{\hat{T}}{2\pi}\sum_{k=-\infty}^{\infty}\int_{-\infty}^{\infty} dq_1 \left(q_1^2+\left(\frac{2\pi k}{L}+A_0\right)^2+m^2\right)^{-s}\,.
\label{zetaapp}
\eeq
Gauge invariance requires invariance under transformation
$A_0\rightarrow A_0 +2\pi k_0/L$, where $k_0$ is integer. This is manifest in (A.3)
since the shift can be absorbed in the sum. We always can look for
a solution for $A_0$ in the interval $|A_0|<\pi/L$, say $A_0=0$.

To evaluate the expression in (\ref{zetaapp}) we will need the following identities
\beqn
&& \Gamma(Z)=\int_0^\infty dt\,t^{z-1}\,e^{-t}\,,
\label{gammaapp}
\\[3mm]
&& \int_0^{\infty}dx (x^2)^{(\alpha-1)/2}(x^2+A^2)^{\beta-1}=\frac{1}{2}(A^2)^{\beta-1+\alpha/2}B(\alpha/2,1-\beta-\alpha/2)\, ,\nonumber\\ [2mm]
&& B(x,y)=\frac{\Gamma(x)\Gamma(y)}{\Gamma(x+y)}\,.
\label{intapp}
\eeqn
The definition of the modified Bessel functions of second kind is
\beq
\int_0^\infty dx\,x^{\nu-1}\,\mbox{exp}\left(-\frac{a}{x}-bx\right)=2\left(\frac{a}{b}\right)^{\nu/2} K_\nu\left(2\sqrt{ab}\right)\,.
\label{mbfapp}
\eeq
The definition of the theta function (see Chapter 21 of \cite{Whitt}) is
\beq
\Theta_3(x,\tau)=\sum_{k=-\infty}^{\infty}q^{k^2}\,e^{2\pi i x}=1+2\sum_{k=1}^\infty\,q^{k^2}\cos{2kx}\,,\quad q=e^{\pi i \tau}\,,
\label{thetaapp}
\eeq
Its Jacobi transformation is
\beq
\Theta_3(x, \tau)=(-i\tau)^{-1/2}\;\mbox{exp}\left(\frac{x^2}{i\pi\tau}\right)\Theta_3(x/\tau , -1/\tau )\,.
\label{jacapp}
\eeq
The evaluation of the zeta function, Eq. (\ref{zetaapp}), proceeds as follows:
\beqn
\zeta(s)&\stackrel{(\ref{intapp})}{=}&\frac{\hat{T}}{2\pi}\frac{\Gamma(\frac{1}{2})\Gamma(s-\frac{1}{2})}{\Gamma(s)}\sum_{k=-\infty}^{\infty}\left[\left(\frac{2\pi k}{L}+A_0\right)^2+m^2\right]^{1/2-s}\,\nonumber\\ [3mm]
&=&\frac{\hat{T}}{2\pi}\frac{\Gamma(\frac{1}{2})\Gamma(s-\frac{1}{2})}{\Gamma(s)}\left(\frac{2\pi}{L}\right)^{1-2s}\sum_{k=-\infty}^{\infty}\left[\left(k+\frac{LA_0}{2\pi}\right)^2+\epsilon^2\right]^{1/2-s}\,\nonumber\\ [3mm]
&\stackrel{(\ref{gammaapp})}{=}&\frac{\hat{T}}{2\pi}\frac{\Gamma(\frac{1}{2})\Gamma(s-\frac{1}{2})}{\Gamma(s)}\left(\frac{2\pi}{L}\right)^{1-2s}\frac{1}{\Gamma(z)}
\nonumber\\ [3mm]
&\times&
\int_0^{\infty}dt\,t^{z-1}e^{-t\alpha^2}\sum_{k=-\infty}^{\infty}e^{-k^2t-k\beta^2 t}\,\nonumber\\ [3mm]
&\stackrel{(\ref{thetaapp})}{=}&\frac{\hat{T}}{2\pi}\frac{\Gamma(\frac{1}{2})\Gamma(s-\frac{1}{2})}{\Gamma(s)}\left(\frac{2\pi}{L}\right)^{1-2s}\frac{1}{\Gamma(z)}
\nonumber\\ [3mm]
&\times&
\int_0^{\infty}dt\,t^{z-1}e^{-t\alpha^2}\Theta_3\left(\frac{i\beta^2 t}{2}, \frac{it}{\pi} \right)\,\nonumber\\ [3mm]
&\stackrel{(\ref{jacapp}),(\ref{thetaapp})}{=}&F\frac{\sqrt{\pi}}{\Gamma(z)}\int_0^{\infty}dt\,t^{z-3/2}e^{-t\alpha^2+\beta^4t/4} \left(1+2\sum_{k=1}^{\infty}e^{-\frac{k^2\pi^2}{t}}\cos{\pi k\beta^2}\right)\,\nonumber\\ [3mm]
&\stackrel{(\ref{mbfapp})}{=}&F\frac{\sqrt{\pi}}{\Gamma(z)}\left(\frac{1}{G^2}\right)^{z-\frac{1}{2}}
\nonumber 
\eeqn
\beqn
&\times&
 \left(\Gamma(z-\frac{1}{2})+4\sum_{k=1}^{\infty}(\pi k G)^{z-\frac{1}{2}}K_{z-\frac{1}{2}}(2\pi k G)\cos{\pi k\beta^2}\right)\,\nonumber\\ [3mm]
&\stackrel{(\ref{mbfapp})}{=}&\frac{\hat{T}L}{4\pi}\frac{1}{m^{2s-2}}\left[\frac{1}{s-1}\right.
\nonumber\\ [3mm]
&+&
\left. \frac{4}{\Gamma(s)}\sum_{k=1}^{\infty}\left(\frac{Lmk}{2}\right)^{s-1}K_{s-1}(Lmk)\cos{LA_0k}\right]\,,
\eeqn
where we introduced intermediate notations 
\beq
\epsilon=\frac{Lm}{2\pi}\,,\quad z=s-\frac{1}{2}\,,\quad F=\frac{\hat{T}}{2\pi}\frac{\Gamma(\frac{1}{2})\Gamma(s-\frac{1}{2})}{\Gamma(s)}\left(\frac{2\pi}{L}\right)^{1-2s}\,, 
\eeq
and  
\beq
\quad \alpha^2=\left(\frac{LA_0}{2\pi}\right)^2+\left(\frac{Lm}{2\pi}\right)^2\,, \quad \beta^2=\frac{LA_0}{\pi}\,, \quad G^2=\alpha^2-\beta^4/4\,.
\eeq
To find the derivative of the zeta function we will make use of the following properties of Euler's $\Gamma$ function:
\beq
\Gamma(z+1)=z\Gamma(z)\,,\quad \Gamma(0)=\infty\,.
\eeq
The derivative is evaluated as follows:
\beqn
 \zeta^\prime(s)&=&\frac{\hat{T}L}{4\pi}\Bigg[-\frac{1}{m^{2s-2}}\frac{1}{(s-1)^2}-\frac{2\ln{m}}{m^{2s-2}(s-1)} \nonumber \\ [3mm]
& & -\frac{4\Gamma^\prime(s)}{\Gamma^2(s) m^{2s-2}}\sum_{n=1}^{\infty}\left(\frac{Lmk}{2}\right)^{s-1} K_{s-1}(Lmk) \cos{LA_0k} \Biggr]\Bigg|_{s=0} \nonumber\\ [3mm]
&=&\frac{\hat{T}Lm^2}{4\pi}\left[-1+\ln{m^2}+8\sum_{k=1}^{\infty}\frac{K_{1}(kLm)}{kLm}\cos{LA_0k}\right]
\eeqn
Following \cite{Haw} we can write the generating functional, 
\beq
\ln Z=\frac{1}{2}\zeta^\prime(0)+\frac{1}{2}\ln\mu^2\zeta(0)\,,
\eeq
where a normalization constant $\mu$ has dimension of mass. Renormalizability 
requires  $$\mu=M_{\rm uv}\,.$$ Thus, in terms of the zeta function and its derivative 
the expression for the effective potential becomes
\beq
V=-\frac{N}{\hat{T}}\left(\zeta^\prime(0)+\zeta(0)\ln{M^2_{uv}}\right)-\frac{N}{4\pi}L m^2\ln\frac{M_{\rm uv}^2}{\Lambda^2}\,.
\eeq
Substituting the expressions for the zeta function and its derivative we obtain 
\beq
V=\frac{NL\omega}{4\pi}\left[1-\ln\frac{\omega}{\Lambda_{CP}^2}
-8\sum_{k=1}^{\infty} \frac{K_1(kL\sqrt{\omega})}{kL\sqrt{\omega}}\cos{kLA_0}\right]\,,
\eeq
where we replaced $m^2$ by $\omega$.

\section*{Appendix B: \\
Kinetic term in case of bosonic theory }

\renewcommand{\theequation}{B.\arabic{equation}}
\setcounter{equation}{0}

\begin{figure}[h!]
\centering
\includegraphics[width=13.5cm]{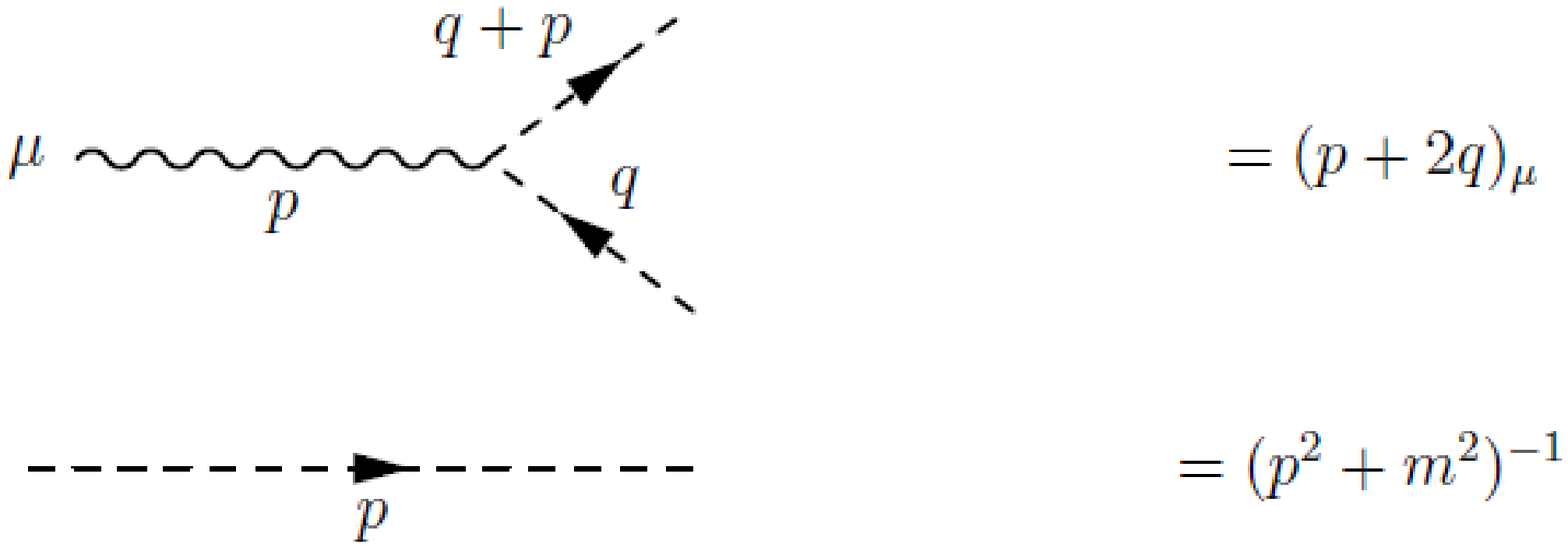}
\caption{\small Feynman rules: vertex and the propagator of $n^l$ field.}
\label{diag2} 
\end{figure}

To find the U(1) charge of the $n^l$ fields one has to consider only the second 
diagram in Fig. (\ref{diag1}). The first diagram is  needed only  for 
renormalization. The relevant part of the action written in the Minkowski spacetime 
takes the form
\beqn
iS_B^M&=&i\int d^2 x \left[\nabla_\mu \bar{n}_l\nabla^\mu n^l-m^2|n|^2\right]\,\nonumber\\ [2mm]
&=&i\int d^2 x \left[\partial_\mu\bar{n}_l\partial^\mu n_l-m^2|n|^2+iA^\mu(\bar{n}_l\overleftrightarrow{\partial}_\mu n^l)+A^2|n|^2\right],
\eeqn
where $\overleftrightarrow{\partial}_\mu=\overrightarrow{\partial}_\mu-\overleftarrow{\partial}_\mu\;$.
We then pass to Euclidean space, 
\begin{center}
$t=-i\tau$\,, \quad $A_0=i\hat{A}_0$\,, \quad $A_i=\hat{A}_i$\,.
\end{center}
The action in  Euclidean space is 
\beq
S_B^E=\int d^2 \hat{x} \left[\partial_k\bar{n}_l\partial_k n_l+m^2|n|^2+i\hat{A}_k
(\bar{n}_l\overleftrightarrow{\partial}_k n^l)+\hat{A}^2|n|^2\right].
\eeq
Now we can determine the Feynman rules. The results are shown in  Fig. (\ref{diag2}).
Thus for the kinetic term (in the case of an infinitely long string) one can write
\beq
\Pi_{ij}=N\int\frac{d^2q}{(2\pi)^2}\frac{(p_i+2q_i)(p_j+2q_j)}{(m^2+q^2)(m^2+(p+q)^2)}\,.
\label{ch0}
\eeq
Introducing the Feynman parameter to combine the denominators
\beq
\frac{1}{\alpha(\alpha+\beta)}=\int_0^1dx\,\frac{1}{(x\beta+\alpha)^2}\,,
\eeq
and substituting $l=q+px$ in Eq. (\ref{ch0}) we arrive at
\beq
\Pi_{ij}=N\int\frac{d^2l\,dx}{(2\pi)^2}\frac{\left[p_ip_j(1-2x)^2-2x(p_il_j+p_jl_i)+4l_il_j\right]}{(l^2+m^2+p^2x(1-x))^2}\,.
\eeq
Terms linear in $l$  vanish. To find the U(1) charge we only need to consider the $p_ip_j$ structure. Thus, the expression for the charge is
\beq
\frac{1}{Ne^2}=\int\frac{d^2l\,dx}{(2\pi)^2}\frac{(1-2x)^2}{(l^2+m^2+p^2x(1-x))^2}=\int_0^1\frac{dx}{4\pi}\,\frac{(1-2x)^2}{m^2+p^2x(1-x)}\,.
\eeq
Expanding the last expression to the zeroth power in $p$ one finally finds
\beq
\frac{1}{Ne^2}=\int_0^1\frac{dx}{4\pi m^2}(1-2x)^2=\frac{1}{12\pi m^2}\,.
\eeq

The case of the  finite length string is considered along similar lines. We recall 
(see \cite{Aff1}) that the limit $p_\mu\rightarrow 0$ is understood as first 
putting $p_0=0$ and then letting $p_1$ go continuously to zero. As a result, 
only $\Pi_{00}\neq 0$. Using the Feynman rules one can derive the following 
expression: 
\beq
\Pi_{00}=\frac{N}{L}\sum_{k=-\infty}^\infty\int\frac{dq}{2\pi}\frac{4\omega_k^2}{(m^2+q^2+\omega_k^2)(m^2+(p+q)^2+\omega_k^2)}\,,
\eeq
where we defined $\omega_k=2\pi k/L$. Introducing again the Feynman parameter and making the same substitution one arrives at
\beq
\Pi_{00}=\sum_{k=-\infty}^\infty\frac{N\omega_k^2}{L}\int_0^1\frac{dx}{(m^2+\omega_k^2+p^2x(1-x))^{3/2}}\,.
\eeq
We expand this expression and keep only the leading power in $p$. Then 
the expression for the charge becomes
\beqn
\frac{1}{Ne^2}&=&\frac{1}{4L}\left[\sum_{k=-\infty}^\infty(m^2+\omega_k^2)^{-3/2}-m^2\sum_{k=-\infty}^\infty(m^2+\omega_k^2)^{-5/2}\right]\nonumber \\ [3mm]
&=&\frac{L^2}{32\pi^3}\left[\sum_{k=-\infty}^\infty(k^2+\alpha^2)^{-3/2}- \alpha^2\sum_{k=-\infty}^\infty(k^2+\alpha^2)^{-5/2}\right]\,,
\label{sumskin}
\eeqn
where $\alpha=Lm/2\pi$. We deal with these sums as follows:
\beqn
S_1(z,\alpha)&\equiv&\sum_{k=-\infty}^\infty(k^2+\alpha^2)^{-z}\stackrel{(\ref{gammaapp})}{=}\frac{1}{\Gamma(z)} \int_0^{\infty}dt\,t^{z-1}e^{-t\alpha^2}\sum_{k=-\infty}^{\infty}e^{-k^2t}\,\nonumber\\ [2mm]
&\stackrel{(\ref{thetaapp})}{=}&\frac{1}{\Gamma(z)} \int_0^{\infty}dt\,t^{z-1}e^{-t\alpha^2}\Theta_3(0,it/\pi)\,\nonumber\\ [2mm]
&\stackrel{(\ref{jacapp})}{=}&\frac{\sqrt{\pi}}{\Gamma(z)} \int_0^{\infty}dt\,t^{z-1}e^{-t\alpha^2}\Theta_3(0,-\pi/it)\,\nonumber\\ [2mm]
&\stackrel{(\ref{mbfapp})}{=}&\frac{\sqrt{\pi}}{\Gamma(z)}\left[\frac{\Gamma(z-\frac{1}{2})}{\alpha^{2z-1}}+4\sum_{k=1}^\infty\left(\frac{k\pi}{\alpha}\right)^{z-\frac{1}{2}}K_{z-\frac{1}{2}}(2k\pi\alpha)\right].
\eeqn

Thus the expression for the charge can be written as
\beqn
\frac{1}{Ne^2}&=&\frac{1}{4L}\left(\frac{L}{2\pi}\right)^3\left[S_1(3/2,\alpha)-\alpha^2 S_1(5/2,\alpha)\right]\,\nonumber\\ [2mm]
&=&\!\!\frac{1}{12\pi m^2}+\frac{L}{2\pi m}\sum_{k=1}^\infty K_1(kLm)\,k-\frac{L^2}{6\pi}\sum_{k=1}^\infty K_2(kLm)\,k^2.
\eeqn
In the limit $Lm\gg1$ the contributions from the modified Bessel functions are exponentially 
small and thus the expression for the charge reduces to that for the  infinitely long string.

\section*{Appendix C: \\
Kinetic term in the supersymmetric case }

\renewcommand{\theequation}{C.\arabic{equation}}
\setcounter{equation}{0}

\begin{figure}[h!]
\includegraphics[width=13.5cm]{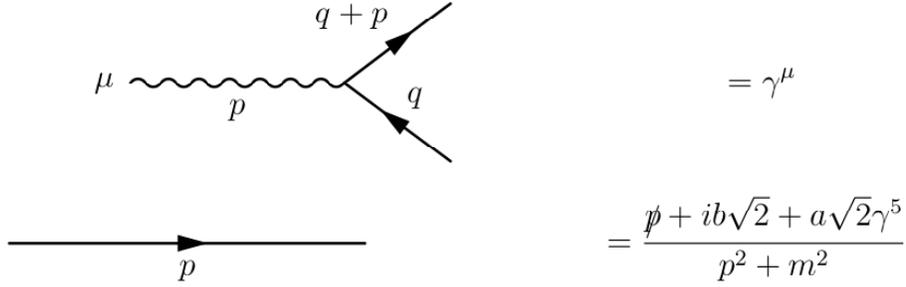}
\caption{\small Feynman rules: vertex and the propagator of $\xi^l$ field.}
\label{fds2} 
\end{figure}

In Appendix B we calculated the first diagram 
(the boson part) in Fig. \ref{fds1}. Now we will calculate the second diagram 
(the fermion part). The relevant part of the fermion action in the Minkowski spacetime is
\beqn
iS_F^M&=&i\int d^2 x \Bigg\lbrace\bar{\xi}\;i\gamma^\mu\nabla_\mu\,\xi
-i\sqrt{2}\sigma\bar{\xi}\left(\frac{1-\gamma^5}{2}\right)\xi \nonumber \\[2mm]
&+&i\sqrt{2}\sigma^*\bar{\xi}\left(\frac{1+\gamma^5}{2}\right)\xi\Bigg\rbrace\,,
\eeqn
where $\nabla_\mu=\partial_\mu-iA_\mu$ is the covariant derivative, and the $\gamma$ 
matrices are defined as  
\begin{center}
\(
\gamma^0=
\begin{pmatrix}
   0 & -i \\
   i & 0
\end{pmatrix}\,,
\quad
\gamma^1=
\begin{pmatrix}
    0 & i\\
    i & 0
\end{pmatrix}\,,
\quad
\gamma^5=
\begin{pmatrix}
    1 & 0 \\
    0 & -1
\end{pmatrix}\,.
\)
\end{center}
We pass to Euclidean space,
\begin{center}
$t=-i\tau$\,, \quad $A_0=i\hat{A}_0$\,, \quad $A_i=\hat{A}_i$\,, \quad 
$\hat{\gamma}^0=\gamma^0$\,, \quad $\hat{\gamma}^1=-i\gamma^1$\,, \quad 
$\hat{\gamma}^5=\gamma^5$\,,
\end{center}
and, since in Euclidean formulation $\xi$ and $\bar{\xi}$ are independent, we define
\begin{center}
 $\hat{\xi}=\xi$\,, \quad $\hat{\bar{\xi}}=i\bar{\xi}$\,.
\end{center}
Thus, the action in Euclidean space can be presented as follows:
\beqn
S_F^E&=&-\int d^2 \hat{x} \Bigg[\hat{\bar{\xi}}\;i\hat{\gamma}^k\hat{\partial}_k
\,\hat{\xi}+\hat{\bar{\xi}}\;\hat{\gamma}^k\hat{A}_k\,\hat{\xi} \nonumber \\[2mm]
&-&\sqrt{2}\sigma\hat{\bar{\xi}}\left(\frac{1-\hat{\gamma}^5}{2}\right)\hat{\xi}
+\sqrt{2}\sigma^*\hat{\bar{\xi}}\left(\frac{1+\hat{\gamma}^5}{2}\right)\hat{\xi}\Bigg]\,.
\label{susyfermacgamma}
\eeqn
Examining this expression in components one can find that it matches that of 
(\ref{susyfermac}). Since from now on all calculations will be carried out in Euclidean 
space we will drop the caret notation. Using (\ref{susyfermacgamma}) we 
find the Feynman rules that are shown in Fig. (\ref{fds2}),
where we introduced a notation $\sigma=a+ib$ and the mass is $m^2=2a^2+2b^2$. 

We begin from the case of the infinitely long string. The fermion
contribution to the kinetic term is 
\beqn
\Pi^{ij}&=&-\int \frac{d^2 q}{(2\pi)^2} \frac{1}{(q^2+m^2)[(p+q)^2+m^2]} \nonumber \\[2mm]
&\times&\Tr\left[\gamma^i(\slashed{q}+
i\sqrt{2}b+\sqrt{2}a\gamma^5)\gamma^j(\slashed{p}+\slashed{q}+
i\sqrt{2}b+\sqrt{2}a\gamma^5)\right]\,.
\eeqn
The Clifford algebra is, as usual,  
\beq
\{\gamma^i\gamma^j\}=2\delta^{ij}\,.
\eeq
As a result, the trace identities for the $\gamma$ matrices become
\beqn
\Tr(\gamma^i\gamma^j)&=&2\delta^{ij}\,,\nonumber\\[1mm]
\Tr(\gamma^i\gamma^j\gamma^k\gamma^l)&=&2\delta^{ij}\delta^{kl}-2\delta^{ik}
\delta^{jl}+2\delta^{il}\delta^{jk}\,,\nonumber \\[1mm]
\Tr(\mbox{odd number of $\gamma$'s})&=&0\,.
\eeqn
Thus, the expression for the kinetic term takes the form
\beqn
\Pi^{ij}&=&-\int \frac{d^2 q}{(2\pi)^2}\frac{\Tr[\gamma^i\slashed{q}\gamma^j
(\slashed{p}+\slashed{q})-m^2\gamma^i\gamma^j]}{(q^2+m^2)[(p+q)^2+m^2]} \nonumber \\[2mm]
&=&-\int \frac{d^2 q}{(2\pi)^2}\frac{1}{(q^2+m^2)[(p+q)^2+m^2]}\nonumber\\[4mm]
&\times&[2q^i(p+q)^j+2q^j(p+q)^i-2q\dot(p+q)\delta^{ij}-2m^2\delta^{ij}]\,.
\label{kinferm}
\eeqn
Notice, that generally speaking $\Tr(\gamma^i\gamma^j\gamma^5)\neq0$ in two 
dimensions. However, we find that both such contributions cancel each other. 

We proceed as in the bosonic theory, introducing the  Feynman parameter and 
making the same substitution. Linear terms drop out, as usual. Furthermore, considering 
only $p^ip^j$ structure we obtain
\beqn
\Pi_F^{ij}&=&p^ip^j\int\frac{d^2 l d x}{(2\pi)^2}\frac{1-(1-2x)^2}
{(l^2+m^2+p^2x(1-x))^2}\nonumber \\[2mm]
&=&p^ip^j\int_0^1 \frac{dx}{4\pi}\frac{1-(1-2x)^2}{m^2+p^2x(1-x)}\,.
\eeqn
Expanding to zeroth order in $p$ we find fermion contribution to $e^2$ ,
\beq
\frac{1}{Ne_F^2}=\frac{1}{6\pi m^2}\,.
\eeq
Combining this with the result we obtained in the boson theory, we finally arrive at
\beq
\frac{1}{Ne^2}=\frac{1}{4\pi m^2}\,.
\eeq

\vspace{2mm}

In the case of the finite length string  the starting expression (\ref{kinferm}) 
is modified
\beqn
\Pi^{ij}&=&-\frac{1}{L}\sum_{k=-\infty}^{\infty}\int\frac{d q}{2\pi}\frac{1}{(q^2+m^2)[(p+q)^2+m^2]}\nonumber\\[2mm]
&\times&[2q^i(p+q)^j+2q^j(p+q)^i-2q\dot(p+q)\delta^{ij}-2m^2\delta^{ij}]\,.
\eeqn
Again, just as in the boson theory we consider $\Pi_{00}$. After we make the 
same substitution and introduce the Feynman parameter we obtain 
\beq
\Pi_{00}=\frac{m^2}{L}\sum_{k=-\infty}^{\infty}\int_0^1 \frac{dx}{(p^2x(1-x)+m^2+\omega_k^2)^{3/2}}\,.
\eeq
Then we expand this expression and keep only the first nonvanishing power in $p$. Thus, 
fermionic contribution to the charge is
\beq
\frac{1}{Ne_F^2}=\frac{m^2}{4L}\sum_{k=-\infty}^\infty(m^2+\omega_k^2)^{-5/2}
\eeq

Summarizing, we obtained a sum identical to that in (\ref{sumskin}). Therefore, their 
evaluation is identical too. Combining the result found in this Appendix with that of the
boson theory, we obtain for the charge
\beq
\frac{1}{Ne^2}=\frac{1}{4\pi m^2}+\frac{L}{2\pi m}\sum_{k=1}^\infty K_1(Lmk)k\,.
\eeq


\newpage 

\begin{thebibliography} {99}
{\small

\bibitem{AharonyKomar}
O.~Aharony and Z.~Komargodski,
{\em The effective theory of long strings}
JHEP {\bf 1305}, 118 (2013)
[arXiv:1302.6257].

\bibitem{ANO}
A.~Abrikosov, Sov.~Phys. JETP {\bf32}, 1442  (1957)
[Reprinted in {\em Solitons and Particles}, Eds. C. Rebbi and G. Soliani
(World Scientific, Singapore, 1984), p. 356];
H.~Nielsen and P.~Olesen, Nucl.~Phys. {\bf B61}, 45 (1973)
[Reprinted in {\em Solitons and Particles}, Eds. C. Rebbi and G. Soliani
(World Scientific, Singapore, 1984), p. 365].

\bibitem{lattice}
A.~Athenodorou, B~ Bringoltz, M.~Teper
{\em Closed Flux Tubes and Their String Description in D=3+1 $SU(N)$ Gauge Theories}
JHEP {\bf 1102}, 030 (2011)
[arXiv:1007.4720 [hep-lat]].


\bibitem{L} 
M.~L\"uscher,
{\em Symmetry Breaking Aspects of the Roughening Transition in Gauge Theories,}
  Nucl.\ Phys.\ B {\bf 180}, 317 (1981).


 \bibitem{HT1}
A.~Hanany and D.~Tong,
{\em Vortices, Instantons and Branes,}
JHEP {\bf 0307}, 037 (2003)
[hep-th/0306150].

\bibitem{ABEKY}
R.~Auzzi, S.~Bolognesi, J.~Evslin, K.~Konishi and A.~Yung,
{\em Non-Abelian superconductors: Vortices and
confinement in N = 2 SQCD,}
Nucl.\ Phys.\ B {\bf 673}, 187 (2003)
[hep-th/0307287].

 \bibitem{SYmon}
{\em Non-Abelian string junctions as confined monopoles,}
Phys.\ Rev.\ D {\bf 70}, 045004 (2004)
[hep-th/0403149].

\bibitem{HT2}
A.~Hanany and D.~Tong,
{\em Vortex Strings and Four-Dimensional Gauge Dynamics,}
JHEP {\bf 0404}, 066 (2004)
[hep-th/0403158].
  
\bibitem{Trev}
D.~Tong, {\em TASI Lectures on Solitons,}
  arXiv:hep-th/0509216.

\bibitem{Jrev}
  M.~Eto, Y.~Isozumi, M.~Nitta, K.~Ohashi and N.~Sakai,
{\em Solitons in the Higgs phase: The moduli matrix approach,}
  J.\ Phys.\ A  {\bf 39}, R315 (2006)
  [arXiv:hep-th/0602170].
  
  \bibitem{SYrev}
M.~Shifman and A.~Yung,
{\sl Supersymmetric Solitons,}
Rev.\ Mod.\ Phys. {\bf 79} 1139 (2007)
[arXiv:hep-th/0703267]; an expanded version in Cambridge University Press, 2009.

\bibitem{Trev2}
D.~Tong,
{\em Quantum Vortex Strings: A Review,}
  Annals Phys.\  {\bf 324}, 30 (2009)
  [arXiv:0809.5060 [hep-th]].

\bibitem{ShYu}
M.~Shifman and A.~Yung,
{\em Non-Abelian strings and the L\"uscher term,}
  Phys.\ Rev.\ D {\bf 77}, 066008 (2008)
  [arXiv:0712.3512 [hep-th]].

\bibitem{W} 
E.~Witten,
{\em Instantons, the Quark Model, and the 1/n Expansion,}
  Nucl.\ Phys.\ B {\bf 149}, 285 (1979).

\bibitem{Aff1}
I.~Affleck,
{\em The Role of Instantons in Scale Invariant Gauge Theories,}
  Nucl.\ Phys.\ B {\bf 162}, 461 (1980).

\bibitem{Laz} 
  G.~Lazarides,
{\em The Effect of Statistical Fluctuations on Confinement and on the Vacuum Structure of the {CP}$^{(n-1)}$ 
Models,}
  Nucl.\ Phys.\ B {\bf 156}, 29 (1979).
  
  \bibitem{Actor}
A.~Actor,
{\em Temperature Dependence of $CP^{N-1}$ Model and the Analogy with Quantum Chromodynamics,}
Fortschr. \ Phys. \ {\bf 33}, 333 (1985).

\bibitem{GSY1} 
A.~Gorsky, M.~Shifman, A.~Yung, {\em Non-Abelian Meissner Effect in Yang-Mills Theories at Weak Coupling,} 
Phys.\ Rev.\ D {\bf 71}, 045010 (2005)
  [hep-th/0412082].

\bibitem{BH}
K.~Bardakci, M.B.~Halpern, {\em Spontaneous Breakdown and Hadronic Symmetries,} Phys. Rev. D 6, 696 (1972)

\bibitem{Abr}
M.~Abramowitz, I.~Stegun, {\sl Handbook of Mathematical Functions: with Formulas, Graphs, and Mathematical Tables}.

\bibitem{GR}
I.S.~Gradshteyn, I.M.~Ryzhik, {\sl Table of Integrals, Series, and Products}. 

\bibitem{SY2} 
  M.~Shifman and A.~Yung, {\em Large-N Solution of the Heterotic N=(0,2) Two-Dimensional CP(N-1) Model,}
  Phys.\ Rev.\ D {\bf 77}, 125017 (2008) (E) Phys.\ Rev.\ D {\bf 81}, 089906 (2010),
  [arXiv:0803.0698 [hep-th]].

\bibitem{C}
S. Coleman, {\em There Are no Goldstone Bosons in Two Dimensions,} Commun. Math. Phys. 31, 259 (1973).

\bibitem{W2}
E. Witten, {\em Theta Dependence in the Large-N Limit of Four-Dimensional Gauge Theories,} 
Phys. Rev. Lett. 81, 2862 (1998). 
 
\bibitem{Whitt}
E.T. Whittaker, G.N. Watson, {\sl A course of modern analysis} (1927).

\bibitem{Haw}
S. W. Hawking, {\em Zeta Function Regularization of Path Integrals in Curved Spacetime,} 
Commun. Math. Phys. 55, 133 (1977).

\bibitem{Milekhin}
A.~Milekhin,
{\em CP(N-1) model on finite interval in the large N limit,}
Phys. \ Rev. D {\bf 86},   105002 (2012)
[arXiv:1207.0417 [hep-th]].

\bibitem{GSYphtr}
A.~Gorsky, M.~Shifman and  A.~Yung,
{\em The Higgs and Coulomb/Confining Phases in ``Twisted-Mass" Deformed CP(N-1) Model,}
Phys.\ Rev.\ D {\bf 73}, 065011 (2006)
  [hep-th/0512153].

\bibitem{Unsal}
G.~V.~Dunne and  M.~\"Unsal,
{\em Resurgence and Trans-series in Quantum Field Theory: The CP(N-1) Model,}
 JHEP {\bf 1211}, 170 (2012) 
[arXiv:1210.2423 [hep-th]].

\bibitem{AdDVecSal}
A.~D'Adda, A.~C.~Davis, P.~DiVeccia and P.~Salamonson,
Nucl.\ Phys.\ {\bf B222} 45 (1983).

\bibitem{W93}
E.~Witten,
  Nucl.\ Phys.\ B {\bf 403}, 159 (1993)
  [hep-th/9301042].
   
\bibitem{ChVa}
S.~Cecotti and C. Vafa,
Comm. \ Math. \ Phys. \ {\bf 158} 569 (1993).


}
  


\end{thebibliography}
\end{document}